\documentclass[preprint,a4paper,showabstract,superscriptaddress]{revtex4-2}
\setcitestyle{super}
\usepackage{hyperref}

\usepackage[utf8]{inputenc}
\usepackage{amsmath,amsfonts,amssymb}
\usepackage{geometry}
\usepackage{graphicx}
\usepackage{color}
\usepackage{xcolor}
\usepackage{empheq}
\usepackage{hyperref}
\usepackage{bm}
\usepackage{url}
\usepackage{diagbox}
\usepackage{array}
\usepackage{setspace}
\usepackage{multirow}
\usepackage{amstext,amsmath,graphicx,bm}

\usepackage{ulem}

\newcommand*{\rev}[1]{\textcolor{black}{#1}}

\def\r'{\mathbf{r'}}

\def\rp{\mathbf{r_p}} 

\begin{abstract}
High-resolution optical imaging of the human retina is essential for diagnosing ophthalmic and neurodegenerative diseases. While adaptive optics combined with coherence gating can resolve retinal structures at the cellular scale, it requires complex, costly, and cumbersome hardware to physically correct for ocular aberrations and motion. Matrix imaging offers a powerful digital alternative by leveraging wave linearity to undo aberrations and multiple scattering{; however, its in vivo deployment has been hindered by} continuous, involuntary eye movements that destroy phase coherence during acquisition. Here, we overcome this fundamental barrier and demonstrate the first in vivo reflection matrix imaging of the human retina using an ultra-simplified optical setup without physical correctors. We introduce a fully computational framework that tracks and compensates for multi-axial ocular motion directly from raw interferometric data, thereby restoring the phase coherence of a multi-spectral distortion matrix. {This enables} local, high-order aberration correction to achieve diffraction-limited, volumetric imaging of retinal tissues. By shifting the hardware burden to {physics-guided numerical computation}, this paradigm opens the way for accessible, non-invasive, and high-resolution digital ophthalmology.

\end{abstract}

\begin{document}

\title{High-resolution retinal matrix imaging via computational motion and aberration correction}

\author         {Aymeric Melt}
\thanks{These authors equally contributed to this work}
\affiliation    {PSL University, ESPCI Paris, CNRS, Institut Langevin, Paris, France}
\author         {L\'{e}o Puyo}
\thanks{These authors equally contributed to this work}
\affiliation    {Institute of Biomedical Optics, University of L\"{u}beck, L\"{u}beck, Germany}
\author         {Victor Barolle}
\affiliation    {OWLO SAS, Paris, France}
\author         {Claude Boccara}
\affiliation    {PSL University, ESPCI Paris, CNRS, Institut Langevin, Paris, France}
\author         {Gereon H\"{u}ttmann}
\affiliation    {Institute of Biomedical Optics, University of L\"{u}beck, L\"{u}beck, Germany}
\affiliation    {Airway Research Center North (ARCN), Member of the German Center for Lung Research (DZL), Germany}
\author         {Alexandre Aubry}
\thanks{Corresponding author: alexandre.aubry@espci.fr}
\affiliation    {PSL University, ESPCI Paris, CNRS, Institut Langevin, Paris, France}
%\email          {alexandre.aubry@espci.fr}

\date{\today}
   \maketitle

   \clearpage 

\noindent {\large \textbf{Introduction}} 

High-resolution optical imaging of the human retina is fundamental for the early diagnosis and monitoring of ophthalmic~\cite{Burns2019,Morgan2022} and neurodegenerative~\cite{London2012,Newman2013} diseases. Over the past decades, Adaptive Optics (AO) combined with Optical Coherence Tomography (OCT) has set the gold standard for resolving retinal structures at the cellular scale~\cite{Hermann2004,Zawadzki2005,Zhang2005,Zhang2006,Torti2009,Kocaoglu2014,Pircher2017,Miller2020,Williams2023,Cai2025}. However, the widespread clinical deployment of AO remains hindered by its substantial hardware complexity, high cost, and limited field of view, requiring real-time physical wavefront sensing and correction. 

A major milestone towards software-driven 3D retinal imaging was achieved a decade ago with computational optical coherence tomography, or holoscopy~\cite{Hillmann2016}. By recording the complex wave field reflected by the eye under a single illumination, holoscopy demonstrated that axial motion~\cite{Hillmann2012} and ocular aberrations~\cite{Hillmann2016} could be corrected a posteriori via digital phase manipulation. {However, relying on \rev{a full-field illumination from a single direction} deprives holoscopy of confocal gating, thereby limiting resolution and contrast due to residual multiple scattering. \rev{Furthermore, aberration compensation in holoscopy relies either on image-metric optimization strategies~\cite{Hillmann2016}, which do not guarantee convergence toward a physically grounded solution, sub-pupil correlation techniques~\cite{Hillmann2019} limited to low-order aberrations, or phase gradient autofocus~\cite{Matkivsky2020} relying on an isoplanatic assumption.}}

Recently, reflection matrix imaging has emerged as a disruptive paradigm to overcome {these limitations}~\cite{kang_high-resolution_2017,badonDistortionMatrixConcept2020,Yoon2020,najarHarnessingForwardMultiple2024,balondradeMultiSpectralReflectionMatrix2024,Zhang2025b}. {By probing the medium with a set of incident plane waves spanning the entire numerical aperture and across multiple wavelengths, matrix imaging acquires the necessary degrees of freedom to synthesize a digital confocal microscope and perform spatially distributed, physics-guided aberration correction. However, translating this matrix paradigm to in vivo retinal imaging has so far remained elusive. Conventional matrix imaging relies on the strict assumption of a static medium during measurement, whereas the living human eye undergoes continuous, involuntary micro-movements (tremors, drifts, and microsaccades) that severely disrupt the phase coherence between successive illuminations.}

{In this work, we overcome this fundamental bottleneck by demonstrating in vivo reflection matrix imaging of the living human retina. Rather than relying on hardware stabilization, we introduce a fully computational framework, inspired by holoscopy~\cite{Hillmann2016}, but extended here to a multi-illumination sequence. It tracks and compensates for multi-axial ocular motion directly from raw experimental data. Once inter-illumination phase coherence is restored, matrix imaging algorithms~\cite{najarHarnessingForwardMultiple2024,balondradeMultiSpectralReflectionMatrix2024} can be fully deployed to correct high-order, non-isoplanatic aberrations across the entire volume, using each retinal layer as an intrinsic distribution of guide stars.}

{As a proof-of-concept, we image a living human retina at \rev{7.5} mm-pupil diameter. We obtain high-contrast images of the nerve fiber layer, inner neural layer with ganglion cells, small vascular structures, and photoreceptor cells with diffraction-limited resolution, while using a drastically simplified optical setup, devoid of any AO components or physical motion trackers. By shifting the burden of aberration and motion correction from hardware to physics-guided numerical computation, this approach paves the way for accessible, widespread high-resolution imaging in clinical ophthalmology.}%

\vspace{5 mm}

\noindent {\large \textbf{Results}}

\noindent {\textbf{Recording the Multi-Spectral Holographic Matrix of a Living Retina}} 

\begin{figure}
    \includegraphics[width=\textwidth]{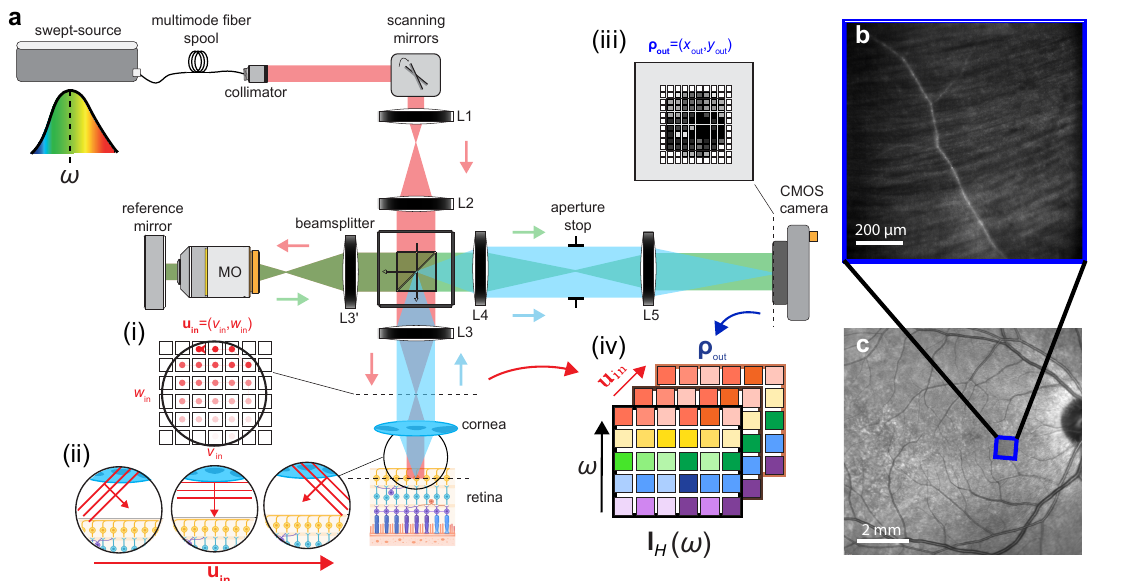}
    \caption{\textbf{Measuring the multi-spectral holographic matrix.} \textbf{a}, A wavelength swept light source illuminates a Linnik{-like} interferometer through a {multi-mode fiber and } collimator, two scanning mirrors and a lens {system (L1-L3)} that allows a raster scanning of the focal spot in the microscope objective (MO) and eye pupil planes ($\mathbf{u}_\textrm{in}$) in each arm (i). The eye, composed of the retina and {the} lens, is placed in the sample arm and is thus illuminated by a set of plane waves at each frequency of the light source bandwidth (ii). The back-scattered wave field is collected through the cornea and the {eye} lens, focused by means of a second lens {system (L4-L6)} on the surface of a CMOS camera where it interferes with a reference beam (iii). The reference beam results from the reflection of the same incident wave-fronts by a reference mirror placed in the focal plane of the MO (NA=0.25). At each frequency $\omega$, for each input wave-front $\mathbf{u}_\textrm{in}$, the interferogram $I_H(\bm{\rho}_\textrm{out},\mathbf{u}_\textrm{in},\omega)$ recorded by each pixel $\bm{\rho}_\textrm{out}$ of the camera provides one column of the response matrix $\mathbf{I}_{H}(\omega)=[I_H(\bm{\rho}_{\textrm{out}},\mathbf{u}_{\textrm{in}},\omega)]$ {(iv)}  {\textbf{b}, Raw OCT image of the nerve fiber layer computed from $\mathbf{I}_{H}(\omega)$ (Methods). \textbf{c}, Scanning laser image of the inspected eye over a wide FOV ($30^\circ \times 30^\circ$, see Methods). The FOV of our own imaging system (\rev{$3.3^\circ \times 3.3^\circ$}) is depicted by a blue rectangle.}}
    \label{fig:setup}
\end{figure}

3D matrix imaging is based on the measurement of a multi-spectral holographic matrix. Inspired from Balondrade \textit{et al.}~\cite{balondradeMultiSpectralReflectionMatrix2024}, the experimental setup and procedure are described in Fig.~\ref{fig:setup} (see Methods). Derived from {holoscopy}~\cite{hillmannHoloscopy2014}, the optical device simply consists in a Linnik{-like} interferometer (Fig.~\ref{fig:setup}a). In the reference arm, a mirror is placed in the focal plane of a microscope objective (MO, NA=0.25). In the sample arm, incident light is focused in the front focal plane of the eye and the retina is illuminated with a collimated beam. This interferometer is illuminated by a swept source laser through: (\textit{i}) a long multimode fiber to create a partial spatial coherence and filter the most diffuse part of the multiple scattering background~\cite{auksoriusMultimodeFiberTool2022, puyoDiffuseilluminationHolographicOptical2023}; (\textit{ii}) two scanning mirrors and a lens that allows a raster scanning of the incident focused beam in front of the MO and eye pupil planes [Fig.~\ref{fig:setup}a(i)]. The retina and reference mirror are thus illuminated by a set of plane waves at each frequency of the light source bandwidth  [Fig.~\ref{fig:setup}a(ii)]. The reflected waves are then collected in the reference and sample arms through the MO and the eye lens, respectively. The two beams ultimately interfere on a camera conjugated with the MO focal plane and the retina. For each input wave-front of frequency $\omega$ and coordinate $\mathbf{u}_\textrm{in}$ in the pupil plane, an interferogram $I_{H}(\bm{\rho}_\textrm{out},\mathbf{u}_\textrm{in},\omega)$ is recorded by the camera whose {pixels} are identified by their conjugated position $\bm{\rho}_\textrm{out}$ in the focal plane [Fig.~\ref{fig:setup}a(iii)]. {$\bm{\rho}_\textrm{out}$ is a lateral vector position in the spatial domain (focal plane), while $\mathbf{u}_\textrm{in}$ maps the spatial frequency domain (pupil plane).}
Each interferogram provides one column of a multi-spectral holographic matrix $\mathbf{I}_{H}(\omega)=\left [I_H(\bm{\rho}_\textrm{out},\mathbf{u}_\textrm{in},\omega) \right ]$ [see Methods and Fig.~\ref{fig:setup}\rev{a}(iv)]. Note that, unlike in our previous work where a similar set up was used to image an opaque ex-vivo cornea~\cite{balondradeMultiSpectralReflectionMatrix2024}, we do not refer yet to this holographic matrix as a reflection (or distortion) matrix because of retinal movement during the acquisition. This will be the case after motion compensation. %The subscripts associated with $\mathbf{A}$ here denote the output and input basis in which the response matrix is defined. Indeed, this matrix will ô  projected between different bases in this work: (\textit{i}) the camera sensor basis ($\mathbf{s}$); (\textit{ii}) the plane wave basis ($\mathbf{u}$) and; (\textit{iii}) the focused basis ($\bm{\rho}$), in which the image is built. The coefficients of the multi-spectral response matrix, $A(\mathbf{s}_\textrm{out},\mathbf{u}_\textrm{in},\omega)$, are noted as functions of the coordinates in the planes associated with each basis. The same convention will be considered for the other operators considered in this paper. 

In the present case, the holographic matrix $\mathbf{I}_H$ is measured with $N_{\textrm{in}} = 112$ plane waves, corresponding to a full scan of the eye pupil. The interferograms are recorded by $N_{\textrm{out}}={{512}^2}$ pixels of the camera, corresponding to an output FOV of 973 $\mu$m (\textit{i.e} $3.3^\circ$ of the visual field), with a spatial sampling ${\delta \rho_0 =\lambda/(2NA)} \sim 1.9$ $\mu$m ($0.006^\circ$). Finally, $N_{\omega}=512$ independent wavelengths are used to probe the retina within the frequency bandwidth of the light source ($800-875$ nm). All the information about the sample is thus contained in the {$15 \times 10^{9}$} coefficients acquired in \rev{1.12} s. In the following, we show how to post-process this wealth of optical data to build a 3D highly-contrasted image of the retina at diffraction-limited resolution despite multi-axial motion, aberration{s} \rev{and residual multiple scattering}. 
\vspace{5mm}

\noindent {\textbf{Holographic OCT and Axial Motion Compensation.}} 

To that aim, the most direct path is to perform, a Fourier transform  over frequency $\omega$ of the back-scattered wave-field recorded for each plane-wave illumination~\cite{povazay_full-field_2006}: This is the principle of full-field swept-source OCT which provides an image whose axial dimension is dictated by photons' times-of-flight. Each resulting image is, however, blurred due to eye axial motion during the wavelength sweep {(Supplementary Figure S1}). The effect of axial motion  causes the apparent optical path lengths in the sample arm to vary with the wavenumber. To circumvent that problem, Hillmann \textit{et al.}~\cite{Hillmann2012} proposed to perform a cross-correlation of sub-bandwidth images for each incident wave $\mathbf{u}_{\textrm{in}}$ to determine the phase fluctuation $\phi(\omega,\mathbf{u}_{\textrm{in}})$ due to axial motion (Methods). This phase error is compensated prior to the calculation of each image:
\begin{equation}
\label{eq1}
    I_G(\bm{\rho}_\textrm{out},z,\mathbf{u}_\textrm{in})=\int_{\omega_-}^{\omega_+} d\omega I_H(\bm{\rho}_\textrm{out},\omega,\mathbf{u}_\textrm{in}) \exp \left [-i \phi(\omega,\mathbf{u}_\textrm{in}) \right] \exp(-2i n \omega z/c ) 
\end{equation}
with $[\omega_{-};\omega_{+}]$, the light source bandwidth, $n$, the eye refractive index, $c$, the vacuum light velocity, $z$, the axial position of the coherence plane defined as the ensemble of points that contribute to the singly-backscattered wave-field at a given time-of-flight. Each resulting matrix, $\mathbf{I}_{G}(\mathbf{u}_\textrm{in})=[I_G(\bm{\rho}_\textrm{out},z ,\mathbf{u}_\textrm{in})]$, provides a three-dimensional image of the retina associated with each incident plane wave $\mathbf{u}_\textrm{in}$ {(Supplementary Figure S1}).

Nevertheless, an axial mismatch subsists between each illumination. The axial shift $\Delta {\hat{z}}(\mathbf{u_{\textrm{in}}})$ of the retina during the illumination sequence can be determined by performing a cross-correlation of each image along the $z$-direction (Methods,Supplementary Figure S2). The holographic images can then be realigned as follows: 
\begin{equation}
\label{eq2}
I_R(\bm{\rho}_\textrm{out},z,\mathbf{u}_\textrm{in})= I_G(\bm{\rho}_\textrm{out},z-\Delta {\hat{z}}(\mathbf{u}_{\textrm{in}}),\mathbf{u}_\textrm{in}).
\end{equation}
The result is a set of three-dimensional images, $\mathbf{I}_R(\mathbf{u}_\textrm{in})=[I_R(\bm{\rho}_\textrm{out},z,\mathbf{u}_\textrm{in})]$, that reveal the main layers of the retina: nerve fiber layer, inner nuclear layer, photoreceptors, retinal pigment epithelium and choroidal tissues {(Supplementary Figure S2)}. However, those images suffer : (\textit{i}) strong aberrations due to defocus and refractive index heterogeneities in the anterior eye segment; (\textit{ii}) multiple scattering {from deep layers} that gives rise to a random speckle background without direct connection with the retina reflectivity. To circumvent those issues, a naive strategy would be to directly sum the realigned images $\mathbf{I}_R(\mathbf{u}_\textrm{in})$ over illumination $\mathbf{u}_\textrm{in}$ in order to synthesize a confocal gate in post-processing and mitigate, at least partially, aberrations and multiple scattering. However, the retina also exhibits a transverse motion during the illumination sequence that should be compensated beforehand. This operation requires a prior compensation of low-order aberrations on each image.
\vspace{5mm}

\noindent {\textbf{Isoplanatic Compensation of Aberrations via Entropy Minimization.}} 

\begin{figure}
\includegraphics[width=\textwidth]{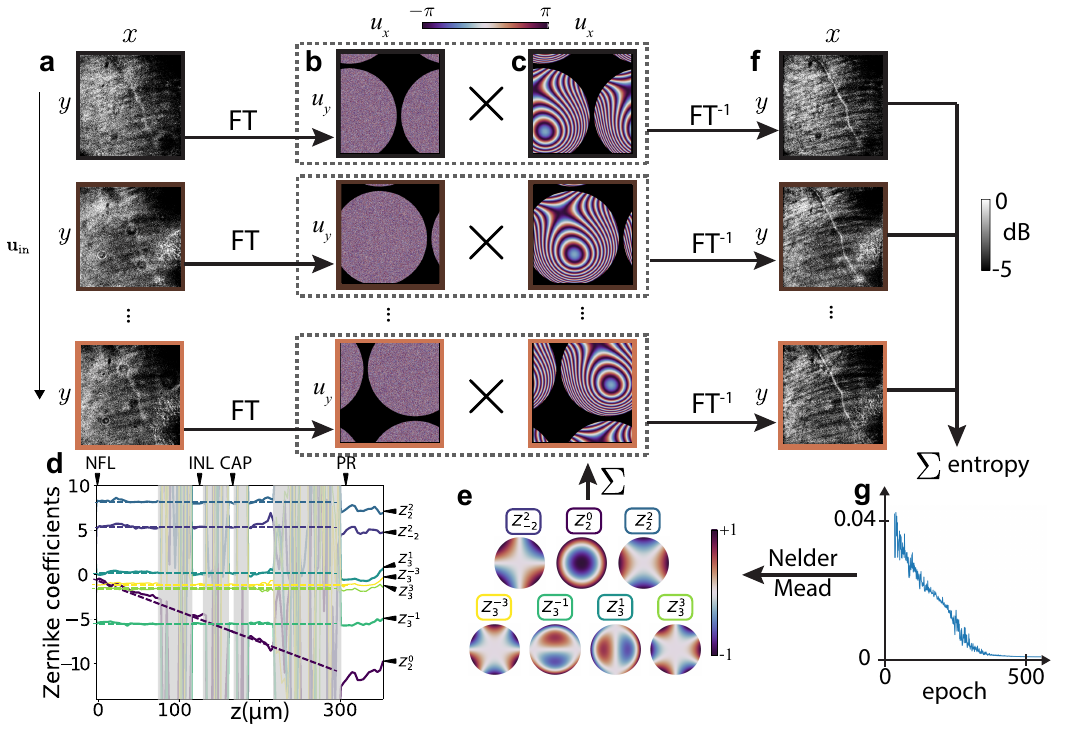}
    \caption{\textbf{Output Aberration Compensation by Entropy Minimization.} \textbf{a}, Set of realigned images, $\mathbf{I}_R(\mathbf{u}_{\textrm{in}})$, recorded for different incident plane waves $\mathbf{u}_{\textrm{in}}$ at the NFL depth. \textbf{b}, Plane wave decomposition of the corresponding images (Eq.~\ref{IH}). \textbf{c}, Element wise product with the de-scan pupil function estimator $\hat{H}_{\text{out}}(\mathbf{u}_{\textrm{out}}-\mathbf{u}_{\textrm{in}},z)$.  {\textbf{d}, Depth evolution of Zernike coefficients $c_n^m(z)$ (continuous lines) associated with the estimator $\hat{H}_{\text{out}}$ (Eq.~\ref{zernike}). The shaded areas correspond to depths where the back-scattered signal is too weak to estimate properly $\hat{H}_{\text{out}}$ and where the coefficients  $c_n^m(z)$ have been interpolated (dashed lines). \textbf{e}, Corresponding Zernike polynomials $Z_n^m(\mathbf{u})$ over the pupil aperture. \textbf{f}, Corrected holographic images $\mathbf{I}_C(\mathbf{u}_{\textrm{in}})$ (Eq.~\ref{correction}). \textbf{g}, Evolution of the loss function through a Nelder Mead iterative procedure that consist in minimizing at each depth $z$ the overall entropy of images, ${I}_C(\bm{\rho}_{\textrm{out}},z,\mathbf{u}_{\textrm{in}})$, obtained for each illumination.}}
    \label{fig:nelder}
\end{figure}

Defocus is related to the axial mismatch between the focal plane of the eye and the coherence plane at each depth $z$. Low-order aberrations are induced by the cornea and eye lens in the vicinity of the pupil plane. Corresponding wave distortions can thus be considered as laterally-invariant {in first approximation}. Under this isoplanatic assumption, each interferometric wave-field $I_R$ can be expressed as follows {(Supplementary Section \rev{S3})}:
\begin{equation}
\label{IH0}
{I}_R(\bm{\rho}_{\textrm{out}},z,\mathbf{u}_{\textrm{in}})=H_{\textrm{in}}(\mathbf{u}_{\textrm{in}},z) \int d\bm{\rho} \, h_{\rev{\textrm{out}}}(\bm{\rho}-\bm{\rho}_{\textrm{out}},z) e^{-j\frac{2\pi}{\lambda f}\mathbf{u}_{\textrm{in}} \cdot (\bm{\rho}-\bm{\rho}_{\textrm{out}})}\gamma(\bm{\rho},z).
\end{equation}
$H_{\textrm{in}}(\mathbf{u}_{\textrm{in}},z) $ is the input transmittance whose support accounts for the pupil size and phase grasps for the wave distortions {accumulated by: (\textit{i}) the incident wave-field between the pupil plane and the coherence plane at depth $z$ in the sample arm; (\textit{ii}) the reference wave-field in the reference arm}. $h_{\textrm{out}}(\bm{\rho}-\bm{\rho}_{\textrm{out}},z) $ is the output point spread function (PSF) connecting the retina ($\bm{\rho}$) at depth $z$  and camera ($\bm{\rho}_{\textrm{out}}$) planes. $\gamma(\bm{\rho},z)$ account for the retina reflectivity at depth $z$. The plane wave decomposition of the realigned wave-field, $$ {I}_R (\mathbf{u}_{\textrm{out}},z,\mathbf{u}_{\textrm{in}})=\sum_{\mathbf{\rho}_{\textrm{out}}} I_R(\mathbf{\rho}_{\textrm{out}},z,\mathbf{u}_{\textrm{in}})\exp \left [-i {2\pi}\mathbf{\rho}_{\textrm{out}}.\mathbf{u}_{\textrm{out}} / (\lambda f) \right ], $$ allows a disentanglement of the retina reflectivity and output aberrations (Eq.~\ref{IH0}):  
\begin{equation}
\label{IH}
 {I}_R(\mathbf{u}_{\textrm{out}},z,\mathbf{u}_{\textrm{in}})=H_{\text{out}}(\textbf{u}_{\text{out}}-\textbf{u}_{\text{in}},z)\Gamma(\textbf{u}_{\text{out}},z)H_{\text{in}}(\textbf{u}_{\text{in}},z)
\end{equation}
where the pupil function $H_{\textrm{out}}$ and the scattering function $\Gamma$ are the Fourier transform of the PSF $h_{\textrm{out}}$ and the reflectivity $\gamma$, respectively.  Figure~\ref{fig:nelder}b illustrates the latter expression by showing how the Fourier spectrum of the retina reflectivity is filtered by the pupil at output and scanned by the tilt of the incident wave-field. {This tilt induces an aliasing phenomenon on each spectrum due to the limited sampling of the holographic wave-field at $\delta \rho_0 \sim \lambda/(2NA)$. This effect could have been mitigated by over-sampling the wave-field at $\lambda/(4NA)$ but this would have been at the cost of a FOV reduction. In the following, we will show how this down-sampling of the holographic wave-field can be maintained in most of the post-processing pipeline.}  

This plane wave decomposition also poses the foundation for an isoplanatic compensation of output aberrations. To do so, it is crucial to estimate the aberration transmittance $H_{\text{out}}$ that we will here express as a linear combination of {second- and third-order} Zernike polynomials~\cite{Mahajan1981} (Fig.~\ref{fig:nelder}d) at each depth :
\begin{align}
\label{zernike}
\hat{H}_{\text{out}}(\mathbf{u},z) =
\sum_{n={2}}^{3}\sum_{m=-n}^{n} c_{n}^m(z) Z_n^m(\mathbf{u}).
\end{align}
Inspired by the seminal work of Hillmann \textit{et al.}~\cite{Hillmann2016}, this estimator is approached by a Nelder Mead optimization procedure~\cite{Nelder1965} (Figs.~\ref{fig:nelder}c-f) that consists in minimizing an entropy-based loss of the corrected images (Fig.~\ref{fig:nelder}f), 
\begin{equation}
\label{correction}
I_C(\bm{\rho}_{\textrm{out}}, z,\mathbf{u}_\textrm{in})= \sum_{\mathbf{u}_{\textrm{out}}} \hat{H}^*_{\text{out}}(\mathbf{u}_{\textrm{out}}-\mathbf{u}_{\textrm{in}}) \mathcal{F}\left \lbrace {I}_R \right \rbrace (\mathbf{u}_{\textrm{out}},z,\mathbf{u}_\textrm{in}) \exp \left [i {2\pi}\mathbf{\rho}_{\textrm{out}}.\mathbf{u}_{\textrm{out}} / (\lambda f) \right ],
\end{equation}
at each depth $z$:
\begin{align}
\label{loss}
\mathcal{L} \lbrace c_n^m(z) \rbrace =\sum_{\mathbf{u}_{\text{in}}} \mathcal{H}\!\left \lbrace \mathbf{I}_{C}(\mathbf{u}_\textrm{in}, z)\right \rbrace,
\end{align}
where $\mathcal{H}$ denotes the Shannon entropy of each {normalized} en-face image intensity {(Methods, Eq.~\ref{entropy})}. Entropy is commonly used as a sharpness metric: sharply focused images concentrate energy in fewer pixels and therefore exhibit lower entropy, whereas blurred images produce more uniform intensity distributions and thus higher entropy. Image entropy is therefore minimized when the phase of the estimator $\hat{H}_{\textrm{out}}$ approaches the isoplanatic aberration transmittance ${H}_{\textrm{out}}$. 

{This process is repeated at each depth $z$ of the retina. Figure~\ref{fig:nelder}d displays the evolution of Zernike coefficients through the different retina layers. A linear evolution of the defocus is recovered until the photoreceptor layer. The other Zernike coefficients remain roughly constant across the retina except in shaded areas where the back-scattered signal is too weak to allow a proper estimation of the aberration transmittance ${H}_{\text{out}}(\mathbf{u},z)$. {A linear interpolation of the Zernike coefficients is therefore performed until the photo-receptor layer (dashed lines in Fig~\ref{fig:nelder}d) in order to ensure a continuity of the estimator $\hat{H}_{\text{out}}(\mathbf{u},z)$ as a function of depth.}}

The result of this optimization process can be appreciated by comparing the original (Fig.~\ref{fig:nelder}a) and corrected (Fig.~\ref{fig:nelder}{f}) images shown at the neural fiber layer (NFL) depth. The contrast and resolution are clearly improved by the phase conjugation of the {aliased} aberration \rev{transmittance whose} phase displayed in Fig.~\ref{fig:nelder}{c}. In particular, the vessel, that was particularly blurred on initial images (Fig.~\ref{fig:nelder}{a}), now appears sharp and more contrasted with respect to background. This gain in image quality is also quantified by the gradual loss of the image entropy during the optimization procedure (Fig.~\ref{fig:nelder}{g}). 

Nevertheless, a lateral shift subsists between the corrected images obtained for each illumination {(Supplementary Figure S3)}. The compensation of this motion is crucial before the combination of each holographic image. As done for the axial shift, the transverse motion $\Delta \bm{\rho} (\mathbf{u_{\textrm{in}}})$ of the retina during the illumination sequence can be determined by performing a cross-correlation of each image in the transverse plane (Methods). The corrected images can then be realigned as follows: 
\begin{equation}
    I_D(\bm{\rho}_\textrm{out},z,\mathbf{u}_\textrm{in})= I_C(\bm{\rho}_\textrm{out}-\Delta \hat{\bm{\rho}}  (\mathbf{u_{\textrm{in}}}),z,\mathbf{u}_\textrm{in}).
\end{equation}

\begin{figure}
    \includegraphics[width=\textwidth]{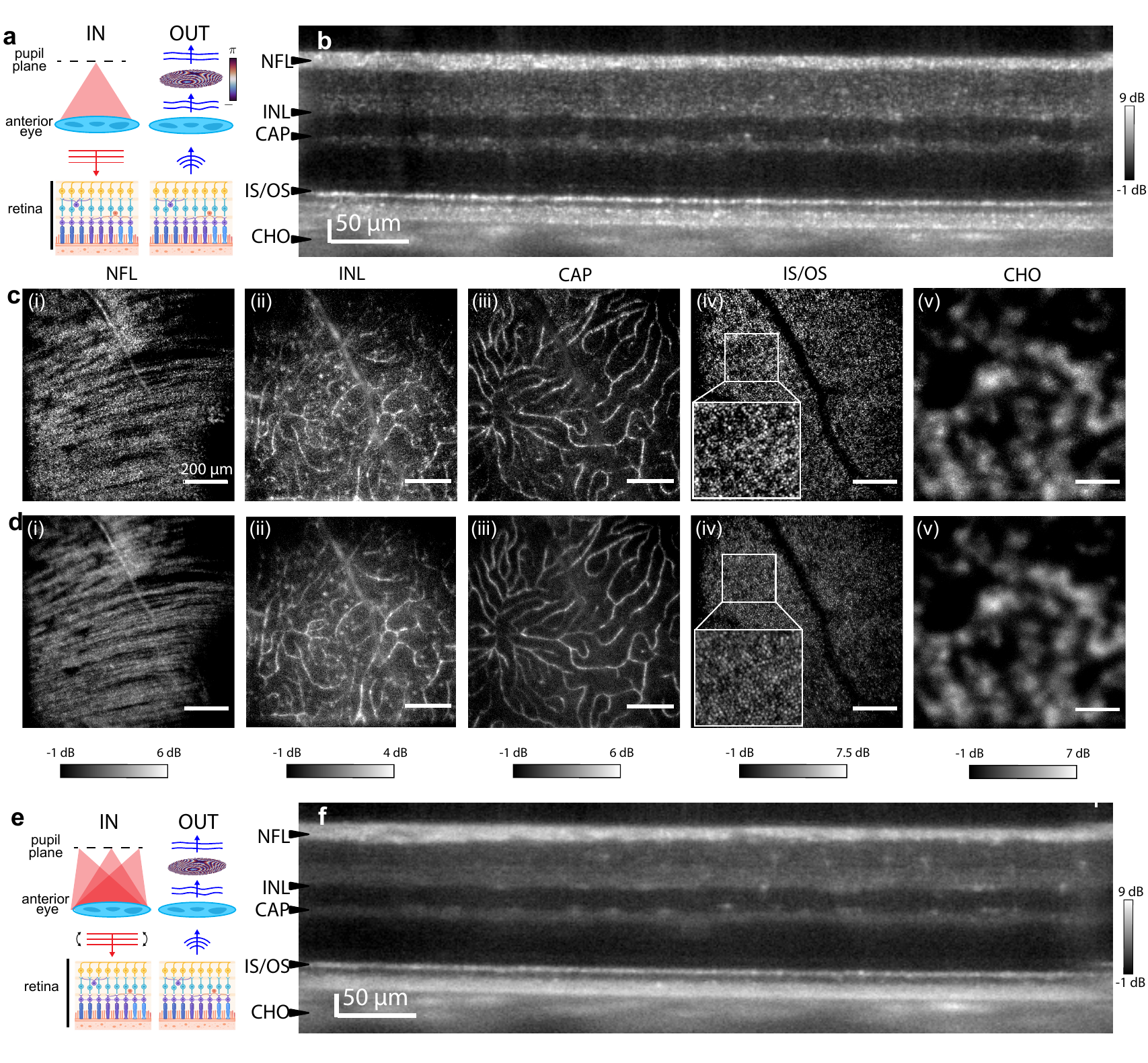}
    \caption{\textbf{Incoherent compound \rev{of plane wave illuminations}}. \rev{\textbf{a} Sketch of holographic OCT under single plane wave illumination and computational aberration compensation~\cite{Hillmann2016}. \textbf{b} Longitudinal section of the corresponding volumetric image after axial motion and output aberration compensation. \textbf{c} Transverse section of the same image at different depths indicated in panel b : (i) Nerve fiber layer; (ii) inner neural layer; (iii) retinal capillaries; (iv) IS/OS junction and (v) underlying choroidal tissue. \textbf{d} Transverse sections of the retina at the same depths after incoherent compound of images obtained from multiple wave illuminations.  \textbf{e} Sketch of the multiple plane wave illumination sequence and isoplanatic aberration compensation at output. \textbf{f} Longitudinal section of the corresponding volumetric image at the same position as panel b.}
    }
    \label{fig:result_bucket}
\end{figure}
\vspace{5 mm}

\noindent {\textbf{Incoherent Compound of Holographic Images.}} 

Now that global retina motion has been compensated, {an} incoherent compound {of holographic images} can be performed to mitigate speckle noise~\cite{puyoDiffuseilluminationHolographicOptical2023}, such that:
\begin{align}
\label{incoherent}
    \mathcal{I}_{I}(\bm{\rho}_{\text{out}},z) = \sum_{\textbf{u}_{\text{in}}}|I_D(\bm{\rho}_{\text{out}},\textbf{u}_{\text{in}}, z)|^2
\end{align}
One longitudinal section of the resulting image $\mathcal{I}_{I}$ is displayed in Fig.~\ref{fig:result_bucket}f. For sake of comparison with standard holoscopy~\cite{Hillmann2016}, the same retina section is also shown in Fig.~\ref{fig:result_bucket}b for a single illumination sequence (Fig.~\ref{fig:result_bucket}a) acquired over an equivalent integration time (Methods). {The multi-illumination sequence (Fig.~\ref{fig:result_bucket}e) provides a smoother image of each retina layer (Fig.~\ref{fig:result_bucket}d) than the single illumination sequence (Fig.~\ref{fig:result_bucket}c), as highlighted by en-face images of the neural fiber layer, Fig.~\ref{fig:result_bucket}c-d(i)], inner neural layer [INL, Fig.~\ref{fig:result_bucket}c-d(ii)] and capillaries [CAP, Fig.~\ref{fig:result_bucket}c-d(iii)]. Such a smooting process can be of interest for highlighting weakly back-scattering components such as ganglion cells [Figs.~\ref{fig:result_bucket}c-d(ii)] or dynamic structures such as capillaries [Figs.~\ref{fig:result_bucket}c-d(iii)], a feature that can be crucial for the early diagnosis of diseases such as glaucoma or neurovascular dysfunctions~\cite{Feng2023,CabreraDeBuc2017}. At larger depth, typically at the inner segment outer segment junction (IS/OS) of photoreceptors, the benefit of incoherent compounding is more significant as the flattening of the multiple scattering speckle brings out the singly-scattered signal from photoreceptors [Figs.~\ref{fig:result_bucket}c-d(iv)]. However, the image contrast and resolution remain largely perfectible. The image resolution, that is here evaluated by the FWHM of its auto-correlation, does not reach the diffraction limit: $\delta \rho\sim$3.6 $\mu$m $> \delta \rho_0$ at the IS/OS junction. The image contrast, that we define here as $\mathcal{C}_I = \mbox{std} [\mathcal{I}_I]/\langle \mathcal{I}_I \rangle$, is also far from being optimal: $\mathcal{C}_I=0.61$ at the IS/OS junction. \rev{The manifestation of multiple scattering is more dramatic in choroid tissues since the image is fully blurred whatever the illumination sequence [Figs.~\ref{fig:result_bucket}c-d(v)]. Blood flow decorrelates so rapidly the holographic wave-field in those highly vascularized tissues that a multi-illumination sequence is useless for incoherent compounding.}

\vspace{5mm}

\noindent {\textbf{Coherent Compound of Holographic Images.}} 

\begin{figure}
    \includegraphics[width=\textwidth]{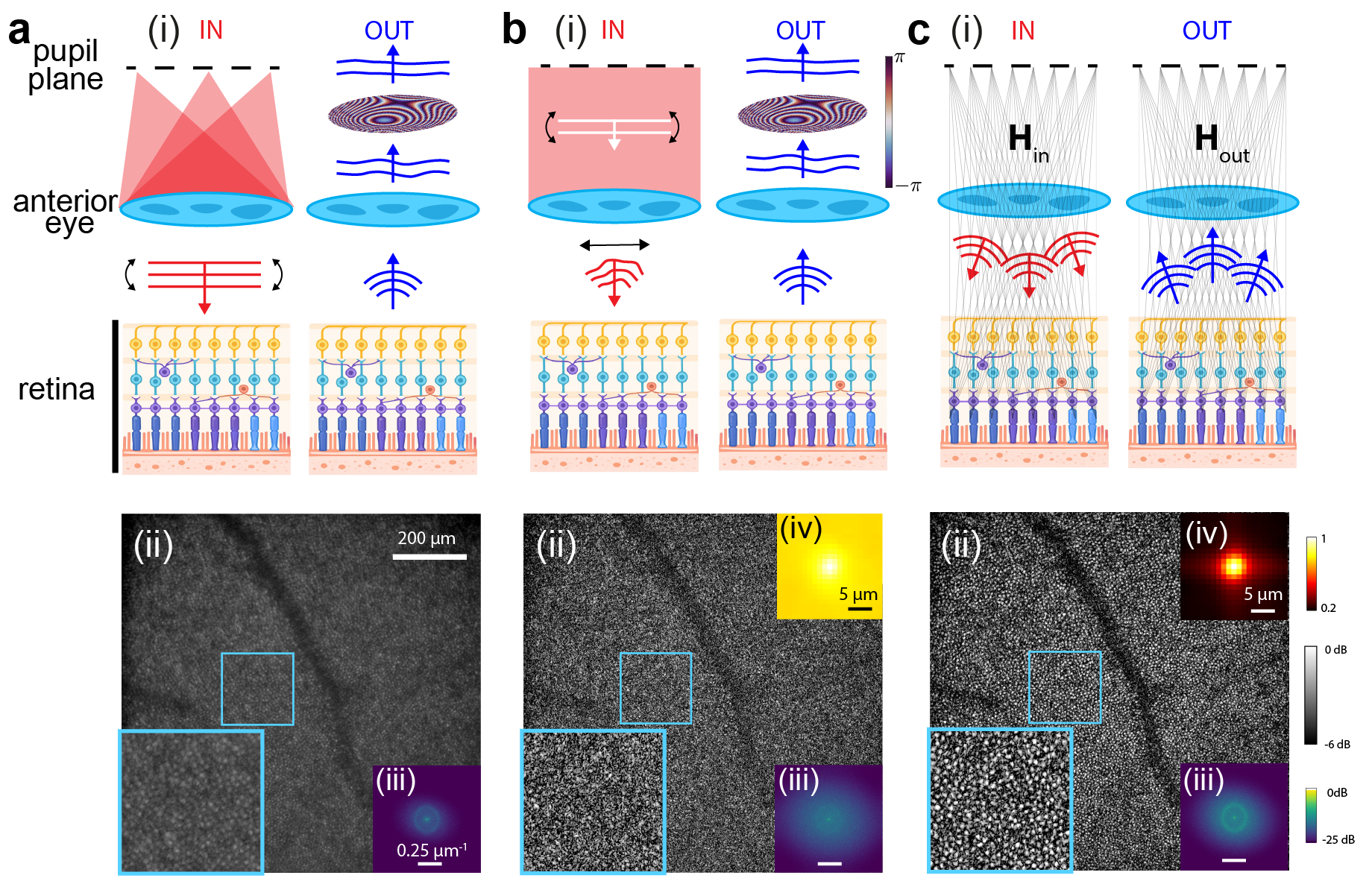}
    \caption{{\textbf{From incoherent to coherent compound of multiple illuminations for imaging photoreceptor cells}. \textbf{a} Incoherent compound imaging (Eq.~\ref{incoherent}). \textbf{b} Coherent compound imaging (Eq.~\ref{IC}). \textbf{c} Confocal image after matrix compensation of input and output aberrations. Each panel contains: (i), a scheme depicting the virtual illumination scheme (left) and aberration correction process applied to the reflected wave-field (right); (ii) transverse cross-section at the IS/OS depth with a zoom as an inset on a sub-part of the FOV displayed as a blue rectangle; (iii) Spatial Fourier transform of each image. (iv) Normalized \rev{reflection point spread function (RPSF, Eq.~\ref{RPSF})} averaged over the whole FOV.}}
    \label{fig:result_correction_rtp}
\end{figure}
The usual way to {reduce the multiple scattering} cross-talk is to perform a confocal gating of back-scattered photons. In the present case, this operation can be performed in post-processing via a coherent sum of each pre-stacked image $I_D$:
\begin{align}
\label{IC}
    \mathcal{I}_{C}(\bm{\rho}_{\text{out}},\rev{z}) = \left | \sum_{\textbf{u}_{\text{in}}} I_D(\bm{\rho}_{\text{out}},\textbf{u}_{\text{in}}, z) \right |^2 .
\end{align}
The resulting image is displayed in Fig.~\ref{fig:result_correction_rtp}{b(ii)} at the IS/OS junction. {Unlike} its incoherent counterpart [Fig.~\ref{fig:result_correction_rtp}{a(ii)}], a direct coherent compound does not reveal the presence of photoreceptors. {This failure is confirmed by the Fourier spectrum of each image displayed in Figs.~\ref{fig:result_correction_rtp}{a-b(iii)}. Although the coherent compound image is characterized by a larger spatial frequency support, the quasi-periodic arrangement of photoreceptors that should manifest as a ring in the spatial Fourier domain is not observed. This destructive coherent compound} can have different origins: (i) low-order input aberrations induced by defocus and anterior eye heterogeneities; (ii) high-order aberrations induced by lateral variations of refractive index exhibited by upstream retina layers; (iii) a residual sub-resolved axial motion that could not have been detected by the cross-correlation between intensity images. 

\begin{figure}
    \includegraphics[width=\textwidth]{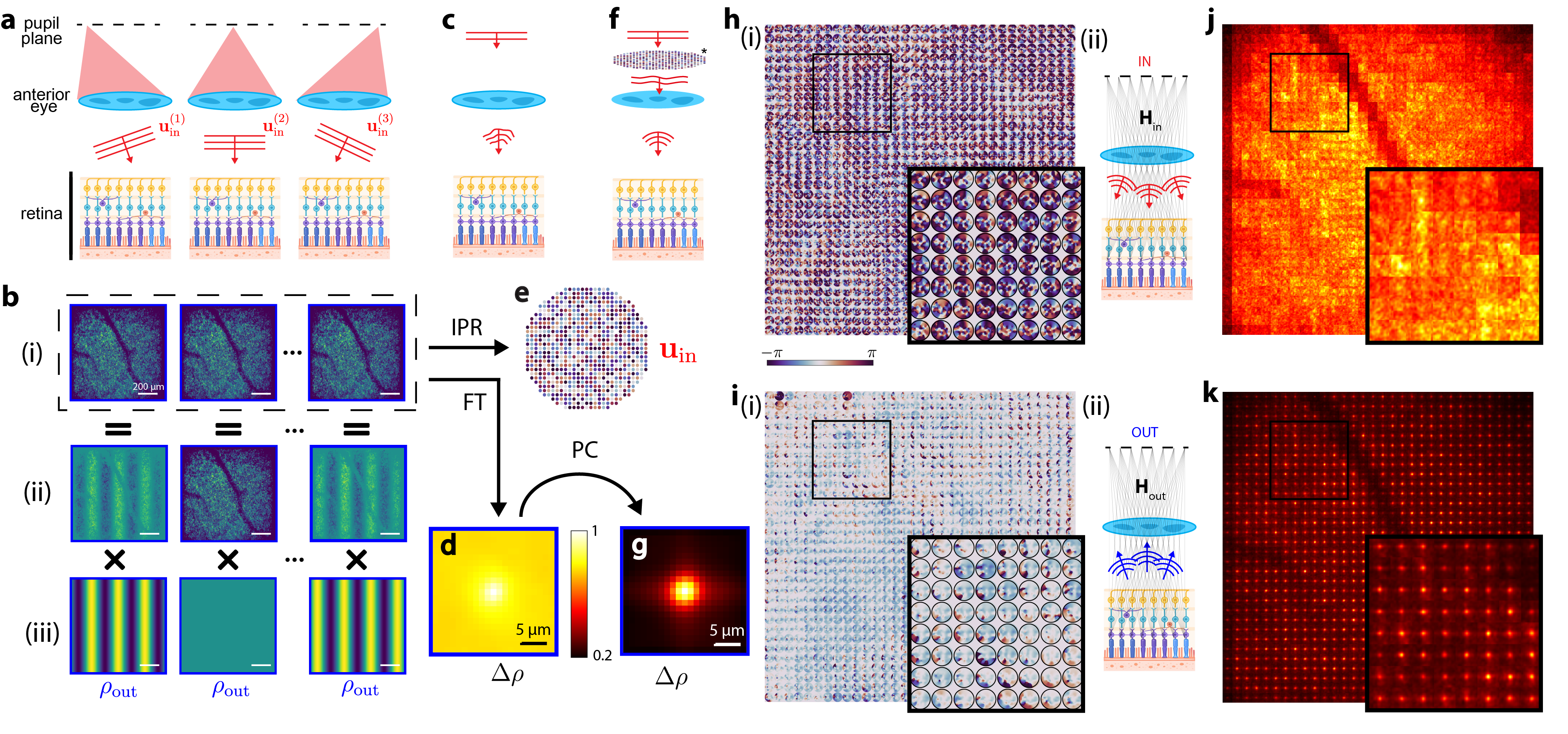}
    \caption{\textbf{Matrix imaging of in-vivo retina}.  \textbf{a} The acquisition sequence consists in illuminating the retina by a set of incident plane waves.  \textbf{b} For each illumination, the interferogram recorded by the camera (i) results from the interference between the wave-field scattered by the retina (ii) and a reference wave-field (iii) induced by a mirror placed in the focal plane of a microscope objective.  \textbf{c} Despite prior motion compensation, coherent compound of each interferogram is affected by aberration resulting in a degraded focal spot.  \textbf{d} These undesirable phenomena can be quantified by performing a Fourier transform of the response matrix over illumination: The resulting point spread function displays a largely predominant diffuse background.  \textbf{e} An iterative phase reversal process applied to the hologram matrix yields an estimator of the aberration transmittance $H_{\textrm{in}}$.  \textbf{f} Its phase conjugate enables a compensation of input aberrations, allowing a coherent combination of input plane waves to produce a diffraction-limited focal spot. \textbf{g} The associated RPSF now displays a diffraction-limited confocal spot on top of a weaker diffuse background.  {\textbf{h} The aberration correction process is repeated at each depth over a set of patches covering the field-of-view, yielding an estimator of the input aberration matrix $\mathbf{H}_{\textrm{in}}$ (i) whose phase conjugate enables a local compensation of aberrations at input of the imaging system (ii). \textbf{i} Same as in h at output of the imaging system. \textbf{j} Initial map of RPSFs across the FOV. Each RPSF is shown over a de-scan area
    of $25\times25$ $\mu$m$^2$. \textbf{k} Final map of RPSFs after matrix compensation of aberrations. Each panel is illustrated by experimental results obtained at the IS/OS junction depth.}}
    \label{fig:fig_rpti_21_04_26_draft}
\end{figure}
All these phenomena manifest as a phase modulation of the transmittance $H_{\textrm{in}}$ in Eq.~\ref{IH0}. If we assume, in first approximation, that output aberrations have been perfectly compensated by the Nelder-Mead optimization process ($h_{\textrm{out}}(\bm{\rho}-\bm{\rho}_{\textrm{out}},z)\simeq \delta (\bm{\rho}-\bm{\rho}_{\textrm{out}})$, with $\delta$ the Dirac distribution), Eq.~\ref{IH0} becomes:
\begin{equation}
\label{IDHin}
   I_{D}(\bm{\rho}_{\text{out}},z,\mathbf{u}_{\textrm{in}}) = H_{\textrm{in}}(\mathbf{u}_{\textrm{in}},z)\gamma(\bm{\rho}_{\textrm{out}},z)
\end{equation}
Under this assumption, each pre-stacked wave-field, $I_{D}(\bm{\rho}_{\text{out}},z,\mathbf{u}_{\textrm{in}})$, is therefore an image of retina reflectivity {[Fig.~\ref{fig:fig_rpti_21_04_26_draft}b(i)]} modulated by a random phase transmittance $H_{\textrm{in}}$. This expression explains why their coherent compound (Eq.\ref{IC}) can be drastically degraded by {residual motion and aberrations undergone by each incident wave}. 

To quantify the level of {those} aberrations, a Fourier transform over $\mathbf{u}_\textrm{in}$ of the pre-stacked wave-fields can be performed to image the focusing process (Fig.~\ref{fig:fig_rpti_21_04_26_draft}c):
\begin{equation}
I_{\Delta}(\bm{\rho}_\textrm{out}, \Delta \bm{\rho} ,z) = \sum_{\mathbf{u}_\textrm{in}}  I_D(\bm{\rho}_{\text{out}}, z,\textbf{u}_{\text{in}}) \exp \left [-i {2\pi} \Delta \bm{\rho}.\mathbf{u}_{\textrm{in}} / (\lambda f) \right ]
\end{equation} 
By injecting Eq.~\ref{IDHin} into the last expression, the resulting wave-field $I_{\Delta}$ is actually shown to provide the input PSF $h_{\textrm{in}}$ modulated by the medium reflectivity $\gamma$:
\begin{equation}
\label{IDHin2}
   I_{\Delta}(\bm{\rho}_{\text{out}},\Delta \bm{\rho},z) = h_{\textrm{in}}(\Delta \bm{\rho} )\gamma(\bm{\rho}_{\textrm{out}}).
\end{equation}
To mitigate the blurring effect of retina reflectivity, a spatial average can be performed over the whole field-of-view:
\begin{equation}
\label{RPSF}
    RPSF(\Delta \bm{\rho},z)=\left \langle \left |  I_{\Delta}(\bm{\rho}_{\text{out}},\Delta \bm{\rho},z) \right |^2 \right \rangle_{\bm{\rho}_{\text{out}}} 
\end{equation}
where the symbol $\langle \cdots \rangle $ stands for average over the variable in subscript. The resulting reflection point spread function (RPSF) is displayed in Fig.~\ref{fig:fig_rpti_21_04_26_draft}{d}. It exhibits the following characteristic shape: a weak confocal spot associated with singly-scattered photons on top of a diffuse background. The level of the latter component explains the bad quality of the raw confocal image displayed in Fig.~\ref{fig:result_correction_rtp}{b}. A local average of the focused wave-field allows us to map the variations of the RPSF across the field-of-view (Methods, {Eq.~\ref{RPSFs}}). The local RPSFs {(Fig.~\ref{fig:fig_rpti_21_04_26_draft}j)} are particularly blurred and show a strong variation across the field-of-view, a feature that could result from a random phase shift due to multi-axial motion between each illumination.  
\vspace{5 mm}

\noindent {\textbf{Matrix Compensation of Aberrations.}} 

To compensate for those aberrations, one can first notice that the set of pre-stacked wave-fields [Fig.~\ref{fig:fig_rpti_21_04_26_draft}{b}(i)] forms a distortion matrix, $\mathbf{D}_{\textrm{in}}(z)=[I_D(\bm{\rho}_{\textrm{out}},\mathbf{u}_{\textrm{in}},z)]$. Introduced in a previous work~\cite{badonDistortionMatrixConcept2020}, this operator is defined mathematically as the element wise product between the reflection matrix of the sample and the phase conjugate of a reference matrix that would be obtained in an homogeneous medium. Indeed, this is exactly what our interferometric set up measures (Fig.~\ref{fig:setup}), the interference between the wave-field reflected by the retina [Fig.~\ref{fig:fig_rpti_21_04_26_draft}{b}(ii)] and a reference wave-field produced by the reference mirror [Fig.~\ref{fig:fig_rpti_21_04_26_draft}{b}(iii)]. In Ref.~\cite{badonDistortionMatrixConcept2020}, it was shown that a singular value decomposition of $\mathbf{D}_{\textrm{in}}$ could lead to an estimation of the aberration transmittance. This is exactly the meaning of Eq.~\ref{IDHin} which, if true, would imply a matrix $\mathbf{D}_{\textrm{in}}$ of rank 1 and an input eigenvector equal to $\mathbf{H}_{\textrm{in}}$. 

However, in practice, the assumption leading to Eq.~\ref{IDHin} is too strong (point-like output PSF) and an iterative phase reversal process  should be rather applied to $\mathbf{D}_{\textrm{in}}$ to estimate properly $\mathbf{H}_{\textrm{in}}$~\cite{najarHarnessingForwardMultiple2024} (Fig.~\ref{fig:fig_rpti_21_04_26_draft}e). Moreover, the local RPSFs displayed in Fig.~\ref{fig:fig_rpti_21_04_26_draft}{j} show a strong variation across the field-of-view. This anisoplanatic feature implies a lateral dependence of the input aberration transmittance such that $\mathbf{H}_{\textrm{in}}(z)=[H_{\textrm{in}}(\bm{\rho},\mathbf{u}_{\textrm{in}},z)]$. A local compensation of aberrations is thus required over restricted spatial windows (\rev{$60.8\times 60.8$ $\mu$m$^2$}). To do so, the iterative phase reversal process is applied to a set of local distortion matrices $\mathbf{D}_{\textrm{in}}$ over each of these windows ({Fig.~\ref{fig:fig_rpti_21_04_26_draft}e}, Methods, {Eq.~\ref{IPR}}). 

The result is displayed in Fig.~{\ref{fig:fig_rpti_21_04_26_draft}}{h} at the IS/OS junction. An estimator $\hat{\mathbf{H}}_{\textrm{in}}(z)$ of the aberration matrix ${\mathbf{H}}_{\textrm{in}}(z)$ is obtained at each depth. It contains the set of input aberration phase laws, {$\hat{{H}}_{\textrm{in}}(\textbf{u}_{\text{in}},\bm{\rho}_{\text{out}},z)$}, obtained for each patch mapping the field-of-view. Compared to the isoplanatic aberration law previously determined by an optimization process at output (Fig.~\ref{fig:nelder}), high-spatial frequency components are retrieved at input. These high-order aberrations originate from sub-resolved retina motion during the illumination sequence but also from refractive index heterogeneities caused by the different retina layers upstream. 

The phase conjugate of the each pupil function  {$\hat{{H}}_{\textrm{in}}(\textbf{u}_{\text{in}},\bm{\rho}_{\text{out}},z)$} can then be used to compensate for these detrimental phenomena {(Fig.~\ref{fig:fig_rpti_21_04_26_draft}f)}:
\begin{align}
\label{RC}
    I'_{\Delta}(\bm{\rho}_{\text{out}},\Delta \bm{\rho},z) =  \sum_{\textbf{u}_{\text{in}}} \hat{H}^*_{\textrm{in}}(\textbf{u}_{\text{in}},\bm{\rho}_{\text{out}},z) I_D(\bm{\rho}_{\text{out}},\textbf{u}_{\text{in}}, z)  \exp \left [-i {2\pi} \Delta \bm{\rho}.\mathbf{u}_{\textrm{in}} / (\lambda f) \right ]
\end{align}
Compared to its original counterpart {(Fig.~\ref{fig:fig_rpti_21_04_26_draft}d)}, the {corresponding RPSF (Eq.~\ref{RPSF})} is drastically improved at the IS/OS junction {(Fig.~\ref{fig:fig_rpti_21_04_26_draft}g)}. 
{The confocal-to-background ratio has been enhanced by 9 dB and its FWHM now reaches the diffraction limit.}

However, the result is still perfectible as only a full-field compensation of aberrations has been performed at output. To tackle anisoplanatic aberrations undergone by the wave-field on its return path, the same aberration correction process can be repeated by exchanging input and output to estimate the output transmittance matrix (Methods). {Note that, for this operation, each holographic wave-field $I'_{\Delta}$ should be over-sampled by a factor 2 at input and output in order to match the confocal resolution ($\delta \rho_c \sim \delta \rho_0/2$) and prevent aliasing of holographic wave-field in the output pupil plane (as previously observed in Fig.~\ref{fig:nelder}b).}  An estimator of the residual aberration matrix, $\delta \hat{\mathbf{H}}_{\textrm{out}(z)}=[\delta \hat{{H}}_{\textrm{out}}(\mathbf{u}_{\textrm{out}},\bm{\rho},z)]$, is obtained at each depth. It contains a set of output residual aberration phase laws for each spatial window inside the field-of-view. A subset is shown at the IS/OS junction depth in Fig.~\ref{fig:fig_rpti_21_04_26_draft}i. Compared to the output aberration law determined by the isoplanatic optimization process (Fig.~\ref{fig:nelder}c), high-spatial frequency components are retrieved and originate from refractive index heterogeneities caused by the different retina layers upstream. Those aberrations are usually not addressed by standard AO that generally relies on isoplanicity. 

The matrix compensation of input and output aberrations [Figs.~\ref{fig:fig_rpti_21_04_26_draft}h-i(ii), Methods] leads to local RPSFs {(Eq.~\ref{RPSFs})} that are drastically improved {(Fig.~\ref{fig:fig_rpti_21_04_26_draft}k)} compared to their original counterpart {(Fig.~\ref{fig:fig_rpti_21_04_26_draft}\rev{j})}. {This optimized focusing quality across the FOV leads to a high-resolution image $\mathcal{I}_M$ (Eq.~\ref{IM}) of photoreceptor cells displayed in {Fig.~\ref{fig:result_correction_rtp}c(ii)}. Compared to a raw coherent compound of holographic images $I_D$ [Fig.~\ref{fig:result_correction_rtp}b(ii)], the gain in image quality provided by matrix imaging is highlighted the spatial frequency spectrum of the final image $\mathcal{I}_M$ that now exhibits a ring resulting from the quasi-periodic arrangement of photoreceptor cells [Fig.~\ref{fig:result_correction_rtp}c(iii)]. Compared to incoherent compound [Fig.~\ref{fig:result_correction_rtp}a(ii)], the gain in image quality is also illustrated by its larger support in the spatial frequency domain [Fig.~\ref{fig:result_correction_rtp}c(iii)]. The image resolution now reaches the diffraction limit ($\delta \rho_0\sim$ 1.9$\mu$m) and its contrast, $\mathcal{C}_M=\mbox{std}[\mathcal{I}_M]/\langle \mathcal{I}_M \rangle$, is improved by a factor 2.5 with regards to incoherent compound.}
\vspace{5mm}

\noindent {\textbf{{Time averaging.}}} 

\begin{figure}
    \includegraphics[width=\textwidth]{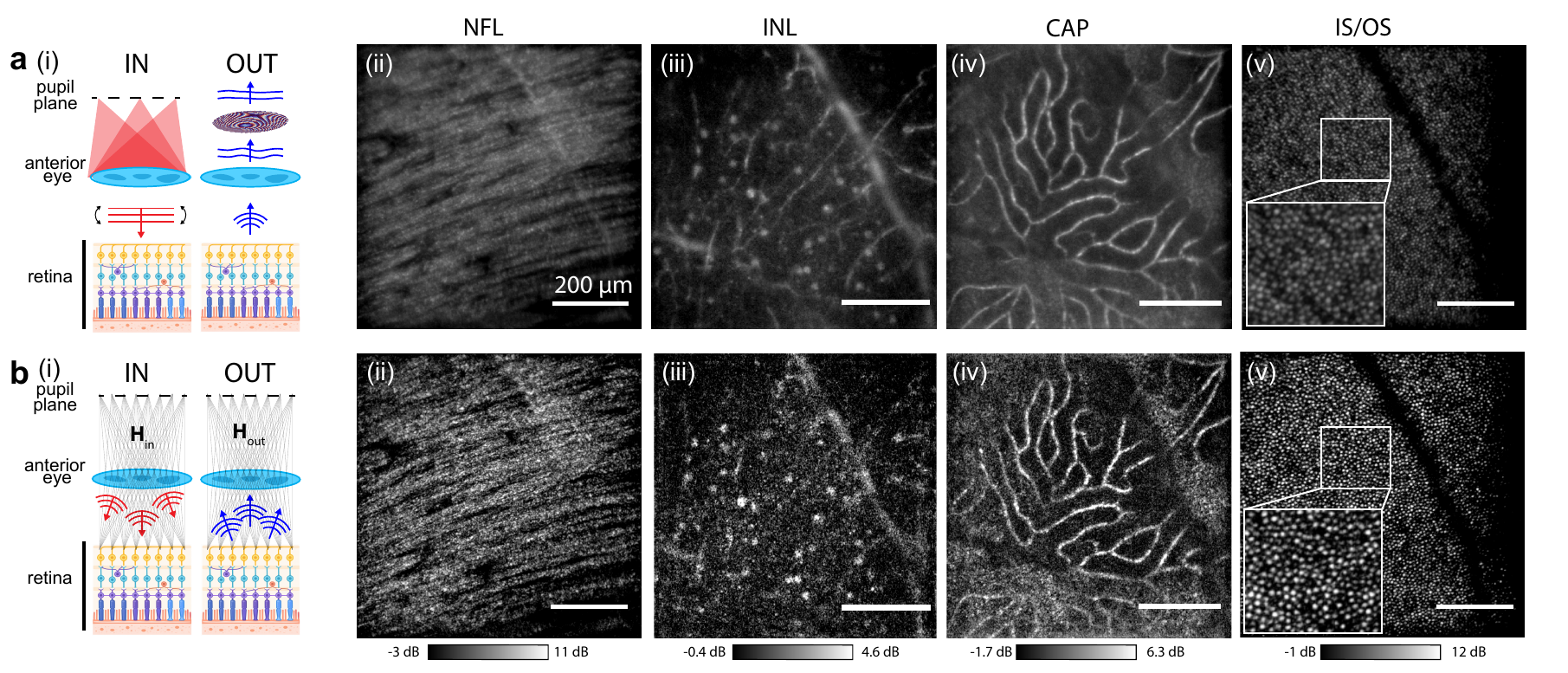}
    \caption{\rev{\textbf{Time averaging of living human retina images}. \textbf{a} Incoherent compound. \textbf{b} Matrix imaging. For each imaging modality, a sketch of the illumination process and post-processing method (i) and transverse sections of the retina are displayed:  NFL (ii), INL (iii), capillaries (iv), IS/OS junction (v).}  Each image is normalized by its spatial average. Each image has been obtained by averaging $\mathcal{I}_I$ and $\mathcal{I}_M$ over a series of 20 multi-spectral holographic matrices (Methods).}
    \label{fig:result_correction_rtp0}
\end{figure}
{Nevertheless, the \rev{shape} of each photoreceptor cell in Fig.~\ref{fig:result_correction_rtp}c(ii) \rev{does not appear perfectly circular}. Those residual distortions are probably due to sub-resolved transverse motion that our post-processing pipeline failed to compensate. To mitigate this undesirable effect, time averaging can be performed (Methods). Figure~\ref{fig:result_correction_rtp0}b shows coherent matrix images averaged over a series of 20 multi-spectral holographic matrices. They are compared to their incoherent counterparts displayed in Figure~\ref{fig:result_correction_rtp0}a.  The result is displayed for the IS/OS segment in Fig.~\ref{fig:result_correction_rtp0}b(v). Each photoreceptor cell now exhibits an almost \rev{circular} shape and the obtained image is equivalent to state-of-the art AO but with a much simpler experimental set up \rev{and a comparable imaging speed~\cite{Kocaoglu2014,Pircher2017} (3.6 MHz voxel rate in the present case)}.}

{The adaptive confocal filter enabled by matrix imaging provides a much better contrast and resolution for photoreceptors [Fig.~\ref{fig:result_correction_rtp0}b(v)] than incoherent compound [Fig.~\ref{fig:result_correction_rtp0}a(v)]. This striking feature holds for the other retina layers displayed in Fig.~\ref{fig:result_correction_rtp0}b. The image of the NFL [Fig.~\ref{fig:result_correction_rtp0}b(ii)] is characterized by a speckle-like feature that we can now attribute to \rev{spatial variations of its reflectivity} rather than to multiply-scattered photons. At the INL inner edge, bright somas are imaged with an optimized contrast thanks to matrix imaging [Fig.~\ref{fig:result_correction_rtp0}b(iii)]. As suggested by a recent AO-OCT study, these somas can be most likely attributed to displaced retinal ganglion cells that appear highly susceptible to glaucomatous damage~\cite{Marte2024}. Finally, time averaging is also crucial to reveal small capillary networks [Fig.~\ref{fig:result_correction_rtp0}b(iv)]. Here, despite blood motion decorrelation, matrix imaging still offers a better resolution and contrast than incoherent compound [Fig.~\ref{fig:result_correction_rtp0}a(iv)].}

\vspace{5 mm}

\vspace{5 mm}

\noindent {\large \textbf{Discussion}}

In this work, we have demonstrated the first in vivo application of reflection matrix imaging to the human retina. While matrix-based approaches have shown remarkable success in static~\cite{kang_high-resolution_2017,badonDistortionMatrixConcept2020,Yoon2020,najarHarnessingForwardMultiple2024,balondradeMultiSpectralReflectionMatrix2024,Zhang2025b} or locally dynamic~\cite{Haim2026} scattering media, translating this formalism to the living eye presents a fundamental challenge due to continuous micro-saccades and ocular drifts. Our results prove that matrix imaging is not restricted to static objects: by tracking and compensating for multi-axial motion, the matrix approach can successfully accommodate dynamic environments. Future work will naturally aim to actively exploit intrinsic tissue motion~\cite{Haim2026}, for instance, by integrating dynamic OCT~\cite{Scholler2020} to refine aberration estimation through local temporal fluctuations

Compared with state-of-the-art high-resolution AO-OCT systems~\cite{Kocaoglu2014,Pircher2017,Miller2020,Williams2023,Cai2025}, our matrix approach achieves an equivalent level of structural detail and spatial resolution with a drastically simplified hardware footprint \rev{and a non-optimized 3.6 MHz voxel rate}. Conventional high-resolution retinal imaging relies heavily on complex wavefront sensors and deformable mirrors, which limit widespread clinical adoption and typically provide only global wavefront correction. In contrast, our framework allows for truly local, high-order aberration correction across the field-of-view.

Beyond simplifying the optical setup, this matrix framework opens up several avenues for deep-tissue ophthalmic imaging. Computationally, it paves the way for advanced strategies like multi-conjugate~\cite{kangTracingMultipleScattering2023,Haim2024} and frequency-dependent~\cite{Lee2023,Giraudat2026} adaptive focusing. These tools will enable complex spatio-temporal focusing laws to enhance OCT penetration depth by realigning multiple scattering paths with the ballistic component. Improved penetration into the choroid layer, in particular, could revolutionize diagnostics for age-related macular degeneration~\cite{Sadeghi2025} and neurodegenerative diseases~\cite{Asanad2019}.   

Finally, this paradigm shift transforms standard OCT into a quantitative imaging modality. By exploiting speckle correlations, it becomes feasible to perform phase-contrast imaging~\cite{Senee2025,Alhaddad2025,McKay2025}, map optical refractive index~\cite{Kandel2019,Hoang2026}, and independently measure scattering parameters across distinct retinal layers~\cite{Hammer1995,Hammer2001,Bocheux2019}. More broadly, the dual ability to dynamically compensate for motion while computationally mitigating multiple scattering opens new frontiers for in vivo exploration of other challenging tissues, such as the skin~\cite{Latriglia2023}, the gastrointestinal tract~\cite{Tsai2014}, or intravascular imaging~\cite{Roleder2015}, in which standard OCT consistently suffers from a severely restricted penetration depth and motion blurring.

\newpage

\noindent {\Large \textbf{Methods}} 

\noindent {\textbf{Experimental components}} 

%The following components were used in the experimental setup (Fig.~\ref{fig:setup}): A swept-source laser (Superlum BS-840-1-OEM with booster; 800-875 nm), one collimator (Thorlabs, F810APC-850), one 300 m long multi-mode fiber (Thorlabs, FG050LGA, low OH, 0.22 NA, 50$\mu$m-core diameter), one pair of galvanometer scanners (Thorlabs, GVS002), one objective (Olympus; Plan Achromat 10$\times$; NA, 0.25), and an ultrafast camera {(60 kHz; Photron, SA-Z)}.
% Light was focused in the front focal plane of the eye, and the retina was illuminated with \textcolor{red}{XX mW} over a field of view of 4 mm diameter.

The experimental setup (Fig.~\ref{fig:setup}) consisted of the following components: a swept-source laser (Superlum BS-840-1-OEM with booster; 800--875 nm), a collimator (Thorlabs, F810APC-850), a 300 m long multimode fiber (Thorlabs, FG050LGA; low OH; NA = 0.22; 50~$\mu$m core diameter), a pair of galvanometer scanners (Thorlabs, GVS002), an objective (Olympus Plan Achromat 10$\times$; NA = 0.25), and an ultrafast camera (Photron SA-Z; 60 kHz).

\vspace{5 mm}

\noindent {\textbf{Ethics and safety}} 

\rev{The study was approved by the ethical board of the University of L\"{u}beck (ethical approval ``Ethik-Kommission L\"{u}beck 16-080''). Investigations were done with one healthy volunteer; written informed consent was obtained from the subject. Compliance with the maximum permissible exposure (MPE) of the retina and all relevant safety rules were confirmed by the responsible safety officer. All experiments were performed in accordance with relevant guidelines and regulations.}
% (ANSI. Z80.36-2021 - light hazard protection for ophthalmic instruments)

\rev{The eye was illuminated with a light power of 3.8~mW over a disc of 4~mm diameter, which corresponds to a power density of 30~mW.cm$^{-2}$. This complies with the ISO 15004-2:2024 standard (``Ophthalmic instruments -- Fundamental requirements and test methods -- Part 2: Light hazard protection''), which permits, for Group 1 (safest) in our configuration (constant illumination for a duration between 25~s and 5.10$^3$~s), 1.0~W.cm$^{-2}$ and 0.7~W.cm$^{-2}$ on the anterior segment and the retina (unweighted), respectively.}

\vspace{5 mm}

\noindent {\textbf{Measurement sampling}} 

Hologram matrices (i.e., OCT volumes) were acquired with $N_{\lambda}=512$ wavelengths spanning the $[800-875]$ nm spectral range of the swept-source laser. The laser operated at a sweep speed of 9050 nm/s, resulting in an inter-volume acquisition time of 10 ms.
The reflection matrix was reconstructed from series of OCT volumes acquired with different illumination angles. Between successive volumes, the illumination angle was varied using the galvanometer scanners and scanned across an input pupil of size \rev{$D_\mathrm{in} \times D_\mathrm{in} = 6.8 \times 6.8$ mm$^2$}.

The acquisition used for the experimental results displayed in Figs.~\ref{fig:setup}, \ref{fig:nelder}, \rev{Figs.~\ref{fig:result_bucket}d-f}, \ref{fig:result_correction_rtp}, \ref{fig:fig_rpti_21_04_26_draft} and Supplementary Figures S1-S3 consisted of $N_u=112$ incident angles, with camera frames of $512 \times 512$ pixels and input wave fields sampled with a spatial pitch of {$\delta u_\mathrm{in}=\rev{510}~\mu$m} in the eye pupil plane. 

The same acquisition was performed for the holographic OCT results displayed in \rev{Figs.~\ref{fig:result_bucket}a-c} but with a galvanometer fixed to normal incidence ($\mathbf{u}_{\textrm{in}}=0$). 

The series of 20 reflection matrix acquisitions used for the experimental results displayed in \rev{Fig.~\ref{fig:result_correction_rtp0}} consisted of $N_u=50$ incident angles, with camera frames of $384 \times 384$ pixels and input wave fields sampled with a spatial pitch of {$\delta u_\mathrm{in}=850~\mu$m} in the eye pupil plane. 

The retina was imaged onto the camera sensor with a magnification factor of 10.4. Given the camera pixel pitch ({${20}$ $\mu$m), an aperture stop was used to limit the detection NA to 0.22 in order to sample interferograms at Shannon-Nyquist limit. Consequently, the output wave field was therefore sampled with a spatial resolution $\delta \rho \sim 1.9$ $\mu$m close to $\lambda/(2\mathrm{NA})$, yielding the following theoretical achievable resolution for a confocal image of the retina: $\delta \rho_c \sim \lambda/(4\mathrm{NA}) \sim {0.9}$ $\mu$m.\\

%Hologram matrices (i.e., OCT volumes) are measured over a number $N_{\lambda}=512$ wavelengths covering the $[800-875]$ nm range of the swept-source laser with a sweep-speed of 9050 nm/s and inter-volume time of 10 ms.
%The reflection matrix is recorded from series of OCT volumes where the illumination angle set by galvanometers is changed in between volumes and scanned over of an input pupil of  $D_\textrm{in} \times D_\textrm{in} = 4 \times 4$ mm$^2$.
%Shorter series are captured with $N_u=112$ incident angles, camera frames of $512 \times 512$ pixel and an input wave-fields spatially sampled at \textcolor{red}{$\delta u_\textrm{in}={307 \mu}$m}.
%Longer series are captured with $N_u=50$ incident angles, camera frames of $384 \times 384 $ and an input wave-fields spatially sampled at \textcolor{red}{$\delta u_\textrm{in}={571 \mu}$m}.
 %The retina was imaged onto the camera sensor with a magnification factor of 10.4. Given the camera pixel pitch ({${20}$ $\mu$m), an aperture stop was used to limit the detection NA to 0.22 in order to sample interferograms at Shannon-Nyquist limit. The output wave field is therefore sampled at a resolution close to $\lambda/(4 \textrm{NA})$, yielding a theoretical achievable resolution for a confocal image of $\delta \rho_c = {1.9}$ $\mu$m.\\
 
 % each wavelength, a number \textcolor{red}{$N_u=112$} of incident plane waves were used to record the reflection matrix.

\vspace{5 mm}

\noindent {\textbf{Data acquisition and graphics processing unit processing}} 
 
% \textcolor{red}{A reprendre par Leo \& Aymeric}

All the interferograms of the \rev{112 plane-wave} acquisition sequence are recorded by the camera in 1.12 s and transferred to the computer for post-processing. \rev{The 20 reflection matrix acquisition takes 10 s.} {The numerical post-processing of the reflection matrix is performed by a graphics processing unit (RTX 6000 Ada Generation). For the data sets considered in this paper, all the focusing and aberration correction algorithms are performed in {15} minutes for the entire volume.}\\

\vspace{5 mm}

\noindent {\textbf{Correction of axial motion during the wavelength sweep of the source}} 

The retina is moving macroscopically during the acquisition of the hologram matrix. This displacement is written as  $\Delta \boldsymbol{r}(\tau) =( \Delta\boldsymbol{\rho}(\tau),\Delta z(\tau)) $. The dependence of this displacement $\Delta \boldsymbol{r}$ with the acquisition time $\tau$ can be rewritten as a function of $\omega $ and $\mathbf{u}_{\textrm{in}}$ since each couple $(\omega,\mathbf{u}_{\textrm{in}})$ corresponds at an acquisition time $\tau$. 

\begin{equation}
    I_F(\bm{\rho}_\textrm{out},\mathbf{u}_{\textrm{in}},z,\omega)=\sum_{\omega_0} I_H(\bm{\rho}_\textrm{out}\mathbf{u}_{\textrm{in}},\omega_0) F(\omega_0 - \omega) e^{-2i n {\omega_0} z/c}
\end{equation}
The width $\delta \lambda$ of each window is of {4.4 nm}. This operation is equivalent in real space to a convolution by an axial window $f(z)$ of extension $\delta z \sim \lambda^2 /(2n_0 \delta \lambda) \sim \rev{60} $ $\mu$m.  The {apparent} axial motion during the wavelength sweep is then obtained from the axial cross-correlation of each filtered image $ I_F(\bm{\rho}_\textrm{out},\mathbf{u}_{\textrm{in}},z,\omega)$ with a reference image at the frequency $\omega_c$ corresponding to $\lambda_c=811$ nm:
\begin{align}
\Delta {z}_{\textrm{opt}} (\omega,\mathbf{u}_{\textrm{in}}) = \underset{\Delta z}{\text{argmax}} \left \lbrace
\left \langle
\left |I_F(\bm{\rho}_{\text{out}}, \mathbf{u}_{\textrm{in}},\omega, z-\Delta z)  I_F(\bm{\rho}_{\text{out}}, \mathbf{u}_{\textrm{in}},\omega_c,z)   \right |^2 \right \rangle_{\bm{\rho}_{\text{out}},z} \right \rbrace \end{align}
The required phase function $\phi(\omega,\mathbf{u}_{\textrm{in}})$ can then be calculated down to a linear term by the integration of $\Delta z_{\textrm{opt}}(\omega,\mathbf{u}_{\textrm{in}})$ over the frequency bandwidth~\cite{Hillmann2012}:
\begin{equation}
\phi(\omega,\mathbf{u}_{\textrm{in}}) = \phi_0(\mathbf{u}_{\textrm{in}}) +  \rev{\frac{2n}{c}}\int_{\omega_-}^{\omega} d\omega  \Delta z_{\textrm{opt}} (\omega,\mathbf{u}_{\textrm{in}}) 
\end{equation}
This phase function $\phi(\omega)$ is then used to compensate axial motion across each wavelength for each plane wave illumination (Eq.~\ref{eq1}). The integration constant $\phi_0(\mathbf{u}_{\textrm{in}})$ defines a phase factor which has no influence on each image intensity, $|I_H(\bm{\rho}_{\text{out}}, \mathbf{u}_{\textrm{in}},\omega_c, z)|^2$, taken independently. However, as illustrated in Fig.~\ref{fig:result_correction_rtp}\rev{b}, it can strongly hamper the raw coherent compound of those images.

\clearpage

\vspace{5 mm}
\noindent {\textbf{Correction of axial motion between each  illumination}}

Beyond this undesirable phase term, the image obtained for each illumination can also exhibit an axial shift with respect to each other. To estimate this displacement, an axial cross-correlation between each intensity image and a reference one {($\mathbf{u}_{\textrm{in}}=\mathbf{0}$)} can be performed:
\begin{align}
\Delta {\hat{z}} (\mathbf{u}_{\textrm{in}}) = \underset{\Delta z}{\text{argmax}} \left \lbrace
\left \langle
\left |I_G(\bm{\rho}_{\text{out}}, \mathbf{u}_{\textrm{in}} , z-\Delta z)  I_G(\bm{\rho}_{\text{out}}, \mathbf{0},z)   \right |^2 \right \rangle_{\bm{\rho}_{\text{out}},z} \right \rbrace \end{align}
{The estimator $\Delta \hat{z} (\mathbf{u}_{\textrm{in}}) $} of the axial displacement is used to realign each image with respect to each other (Eq.~\ref{eq2}).

\vspace{5 mm}
\noindent {\textbf{Correction of transverse motion between each  illumination}}

To estimate the transverse motion between the different volumes after isoplanatic aberration correction via entropy minimization, a similar strategy is followed by computing a transverse cross-correlation between each intensity image and the reference one {($\mathbf{u}_{\textrm{in}}=\mathbf{0}$)}:
\begin{align}
 \Delta {\hat{\boldsymbol{\bm{\rho}}}}(\mathbf{u}_{\textrm{in}}) = \underset{\Delta \bm{\rho}}{\text{argmax}} \left \lbrace
\left \langle
\left |I_C(\bm{\rho}_{\text{out}}-\Delta \bm{\rho}, \mathbf{u}_{\textrm{in}}, z)  I_C(\bm{\rho}_{\text{out}}, \mathbf{0},z)   \right |^2 \right \rangle_{\bm{\rho}_{\text{out}},z} \right \rbrace \end{align}

\vspace{5 mm}
\noindent {\textbf{Loss function}}

The loss function in Eq.~\ref{loss} is computed by first normalizing en-face image intensity, such that:
\begin{equation}
    \mathcal{I}_{\rev{N}}(\bm{\rho}_{\textrm{out}},\mathbf{u}_\textrm{in}, z)=|{I}_C(\bm{\rho}_{\textrm{out}},\mathbf{u}_\textrm{in}, z)|^2/\sum_{\bm{\rho}_{\textrm{out}}}|{I}_C(\bm{\rho}_{\textrm{out}},\mathbf{u}_\textrm{in}, z)|^2.
\end{equation}
The Shannon entropy $\mathcal{H}$ of each en-face image, $\mathbf{I}_{C}(\mathbf{u}_\textrm{in}, z)$, is then given by: 
\begin{equation}
\label{entropy}
\mathcal{H}\!\left \lbrace \mathbf{I}_{C}(\mathbf{u}_\textrm{in}, z) \right \rbrace =- \sum_{\bm{\rho}_{\textrm{out}}} \mathcal{I}_{\rev{N}}(\bm{\rho}_{\textrm{out}},\mathbf{u}_\textrm{in}, z) \log \left [ \mathcal{I}_{\rev{N}}(\bm{\rho}_{\textrm{out}},\mathbf{u}_\textrm{in}, z) \right ] .
\end{equation}

\noindent\textbf{Local input RPSF}

To probe the local RPSF, the field-of-view is divided into regions that are defined by their central midpoint $\mathbf{r}_p = (\rho_p, z)$ and their lateral extension $L$. A local average of the de-scanned intensity, $|I_{\Delta}|^2$, can then be performed in each region:
\begin{equation}
\label{RPSF2}
    RPSF(\Delta \bm{\rho},\mathbf{r}_p)=\left \langle \left |  I_{\Delta}(\bm{\rho}_{\text{out}},\Delta \bm{\rho},z) \right |^2 W_L(\bm{\rho}_{\text{out}}-\bm{\rho}_p)\right \rangle {\bm{\rho}_{\text{out}}} 
\end{equation}
where {$W_{L}(\bm{\rho}_{\text{out}} - \bm{\rho}_\textrm{p}) = 1$ for $|x_{\text{out}} - x_\mathrm{p}|<\rev{L/2}$ and $|y_{\text{out}} - y_\mathrm{p}|<\rev{L/2}$} , and zero otherwise. A delicate compromise has to be made on the size $L$ of the window function: it must be small enough to approach the isoplanatic condition, but large enough to encompass a sufficient number $N_L=(L/\delta{\rho})^2$ of independent realizations in order to smooth the speckle reflectivity (with $\delta \rho \sim \lambda/(2NA) $ the standard diffraction-limited resolution). In the present case, we chose $N_L=64^2=4096$. The expression provided in Eq.~\ref{RPSF2} is used to build the map of RPSF displayed in Fig.~\ref{fig:fig_rpti_21_04_26_draft}{j}. \\

\noindent\textbf{Correction of input aberrations}

As mentioned in the core of the paper, the bad focusing quality exhibited by Figs.~\ref{fig:fig_rpti_21_04_26_draft}\rev{d and j} is the result of of a random phase shift $\phi_0$ we discussed above and of aberrations induced by refractive index heterogeneities both in the anterior eye and across the different retina layers. This latter source of aberration as well as the phase shift are anisoplanatic and requires a local compensation. 

To that aim, aberrations are analyzed locally over the spatial windows $W_L$ discussed above by computing a local correlation matrix of wave distortions around each point $\mathbf{r}_p$ of the field-of-view~\cite{najarHarnessingForwardMultiple2024}: 
\begin{equation}
\label{corr_in}
    C_\textrm{in} (\mathbf{u}_\textrm{in}, \mathbf{u}'_\textrm{in},\mathbf{r}_p)= \left \langle I_D(\mathbf{u}_\textrm{in},{\bm{\rho}_\textrm{out},z})I_D^*(\mathbf{u}'_\textrm{in},{\bm{\rho}_\textrm{out},z}) W_{{L}}({\bm{\rho}_\textrm{out} - \bm{\rho}_\textrm{p}})\right \rangle _{{\bm{\rho}_\textrm{out}}}
\end{equation}
Iterative phase reversal is then applied to each correlation matrix $\mathbf{C}_\textrm{in}(\mathbf{r}_p)$ to retrieve an estimator of the aberration matrix $\hat{\mathbf{H}}_{\textrm{in}}=[H_{\textrm{in}}(\mathbf{u}_\textrm{in},\mathbf{r}_p)]$ (see further). The phase conjugate of $\hat{\mathbf{H}}_{\textrm{in}}$ compensate for the wave distortions undergone by the incident wave-fronts (Eq.~\ref{RC}).
However, the corrected wave-fields ${\mathbf{I}}'_\textrm{in}$ are only intermediate since residual phase distortions undergone by the reflected wave-fronts remain to be corrected.

To that aim, an output de-scanned wave-field  ${I}'(\Delta \bm{\rho}_{\textrm{out}},\bm{\rho}_{\textrm{in}},z )$ is deduced from the input de-scanned wave-fields using the following change of variable~\cite{najarHarnessingForwardMultiple2024}:
\begin{equation}
\label{change}
   I'_{\Delta}(\Delta \bm{\rho}_\textrm{out},\bm{\rho}_\textrm{in},z)=I'_{\Delta}(\bm{\rho}_\textrm{in}+\Delta \bm{\rho}_\textrm{out},-\Delta \bm{\rho}_\textrm{out},z)
\end{equation}
with $\Delta \bm{\rho}_\textrm{out}=\bm{\rho}_\textrm{out}-\bm{\rho}_\textrm{in}=-\Delta \bm{\rho}_\textrm{in}$. A set of output distortion matrices $\mathbf{D}_{\textrm{out}}(z)=[  I'( \mathbf{u}_\textrm{out},\bm{\rho}_\textrm{in},z)]$ is then built at each depth by applying a Fourier transform over the de-scanned coordinate:
\begin{equation}
I'_D(\mathbf{u}_\textrm{out},\bm{\rho}_\textrm{in},z)= \int d\Delta \bm{\rho}_\textrm{out} I'_{\Delta}(\Delta \bm{\rho}_\textrm{out},\bm{\rho}_\textrm{in},z) \exp \left ( -i \frac{2\pi}{\lambda f}{\mathbf{u}_\textrm{out} \cdot \Delta \bm{\rho}_\textrm{out}}\right)
\end{equation}
From {${\mathbf{D}}_{\textrm{out}}$}, one can build a correlation matrix $\mathbf{C}_{\textrm{out}}(\mathbf{r}_p)$ of output wave distortions around each point $(\bm{\rho}_p,z)$ in the field-of-view. Its coefficients write:
\begin{equation}
\label{corr_out}
    C_{\textrm{out}} (\mathbf{u}_{\textrm{out}}, \mathbf{u}'_{\textrm{out}},\mathbf{r}_p)= \left \langle I'_{D}({\bm{\rho}_\textrm{in} },\mathbf{u}_{\textrm{out}},z)I'^*_{D}({\bm{\rho}}_{\textrm{in} },\mathbf{u}'_{\textrm{out}},z) W_{L}({\bm{\rho}_\textrm{in} - \bm{\rho}_\textrm{p}})\right \rangle _{{\bm{\rho}_\textrm{in}}}
\end{equation}
The IPR algorithm described further is then applied to each matrix $\mathbf{C}_{\textrm{out}}(\mathbf{r}_p)$. The resulting output phase laws, $\hat{\bm{\phi}}_\textrm{out}(\mathbf{r}_p)$, are leveraged to compensate for the residual wave distortions undergone by the reflected wave-fronts:
\begin{align}
\label{RC2}
   I''_\Delta(\Delta \bm{\rho}_\textrm{out},\bm{\rho}_\textrm{in},z)=   \sum_{\textbf{u}_{\text{out}}} \hat{H}^*_{\textrm{out}}(\textbf{u}_{\text{out}},\bm{\rho}_{\text{in}},z) I'_D(\textbf{u}_{\text{out}},\bm{\rho}_{\text{in}}, z)  \exp \left [-i {2\pi} \Delta \bm{\rho}_\textrm{out}.\mathbf{u}_{\textrm{in}} / (\lambda f) \right ]
\end{align}
The RPSFs displayed in Figs.~\ref{fig:fig_rpti_21_04_26_draft}{k} are built from these corrected wave-fields:
\begin{equation}
\label{RPSFs}
    RPSF(\Delta \bm{\rho}_{\textrm{out}}\rev{,\mathbf{r}_p})=\left \langle \left |  \rev{I''_{\Delta}}(\Delta \bm{\rho}_{\textrm{out}},\bm{\rho}_{\text{in}},z) \right |^2 W_L(\bm{\rho}_{\text{out}}-\bm{\rho}_p) \right \rangle_{\bm{\rho}_{\rev{\text{in}}}} 
\end{equation}
The matrix images displayed in Fig.~\ref{fig:result_correction_rtp}{c} are correspond to the confocal component of these corrected wave-fields ($\Delta \bm{\rho}_{\textrm{out}}=\mathbf{0}$):
\begin{equation}
\label{IM}
    \mathcal{I}_M(\bm{\rho}_\textrm{in},z)=I''_\Delta(\mathbf{0},\bm{\rho}_\textrm{in},z)
\end{equation}

\noindent {\textbf{Iterative phase reversal algorithm.}} 

The IPR algorithm is a computational process that provides an estimator of the {pupil transmittance matrix}, $H(\mathbf{u},\rp)=\exp \left [ i \phi(\mathbf{u},\mathbf{r}_p) \right ] $, that links each point $\mathbf{u}$ of the pupil plane with each voxel $\mathbf{r}_p$ of the retina volume. To that aim, the correlation matrix $\mathbf{C}$ computed over the spatial window $W_{L}$ centered around each point $\mathbf{r}_p$ is considered {(Eqs.~\ref{corr_in} and ~\ref{corr_out})}.  Mathematically, the algorithm is based on the following recursive relation:
\begin{equation}
\label{IPR}
{\hat{\bm{\phi}}^{(n)} (\mathbf{r}_p) = \mathrm{arg}\left\{  \mathbf{C}(\mathbf{r}_p) \times \exp \left [ i \hat{\bm{\phi}}^{(n-1)}(\mathbf{r}_p) \right ]  \right\}}
\end{equation}
where {$\hat{\bm{\phi}}^{(n)}$} is the estimator of {${\bm{\phi}}$} at the $n^\textrm{th}$ iteration of the phase reversal process. $ \hat{\phi}^{(0)}$ is an arbitrary wave-front that initiates the process (typically a flat phase law) and ${\hat{\bm{\phi}}}= \lim_{n\to\infty} {\hat{\bm{\phi}}}^{(n)}$ is the result of IPR. \\

\noindent\textbf{Data availability.} Optical data used in this paper have been deposited on Zenodo at \href{https://zenodo.org/record/23345183}{https://zenodo.org/record/23345183}(ref.~\cite{puyo2026}).
\\
\vspace{5 mm}

\noindent\textbf{Code availability.}
Codes used to post-process the optical data within this paper are available upon request on Zenodo at \href{https://zenodo.org/records/23345976}{https://zenodo.org/records/23345976}(ref.~\cite{melt2026}).\\
\vspace{5 mm}

\noindent\textbf{Acknowledgments.}
We thank P. M\^{e}c\'{e} and P. Balondrade for fruitful discussions about the experimental results of this study. \\
\vspace{5 mm}

\noindent\textbf{Funding Information.}
We are grateful for the funding provided by the European Research Council (ERC) under the European Union's Horizon 2020 research and innovation programme (grant agreements no. 819261 and 101158062, REMINISCENCE and MUSE projects, A.A.) This project has also received funding from Labex WIFI (Laboratory of Excellence within the French Program Investments for the Future; grant nos ANR-10-LABX-24 and ANR-10-IDEX-0001-02 PSL*), from Agence Nationale de la Recherche (Grant No. ANR-22-ASTR-0020 under the AquaMat project, A.A.), from CNRS Innovation (Prematuration programme, MATRISCOPE project; A.A.), from PSL (Prematuration program, MIRE project) and from the German Science Foundation (DFG, HU 629-6, GH).\\
\vspace{5 mm}

\noindent\textbf{Author Contributions.}
V.B., L.P., G.H. and A.A. initiated the project. V.B. and A.A. supervised the project. L.P., V.B. and G.H. designed the experimental setup. L.P. built the experimental set up. A.M., L.P. and V.B. designed the acquisition scheme. A.M. and V.B. developed the post-processing tools. L.P. performed the living human retinal imaging experiment. A.M., L.P., V.B. and A.A. analyzed the experimental results. A.M., L.P. and A.A. prepared the manuscript. A.M., L.P., V.B., C.B., G.H. and A.A. discussed the results and contributed to finalizing the manuscript. \\
\vspace{5 mm}

\noindent\textbf{Competing interests.}
V.B., C.B. and A.A. are inventors on a patent related to
this work held by CNRS (no. US11408723, published August 2022). V.B. and A.A. are co-founders of the OWLO company, that will commercialize the microscope used in this study. The other authors declare no competing interests.

\clearpage

\def\bibsection{\noindent \textbf{References.}}

\bibliography{MyLibrary.bib}

\clearpage

\renewcommand{\thetable}{\arabic{table}}
\renewcommand{\thefigure}{\arabic{figure}}
\renewcommand{\theequation}{S\arabic{equation}}
\renewcommand{\thesection}{\arabic{section}}
\renewcommand{\thesubsection}{\arabic{section}.\arabic{subsection}}
\renewcommand{\figurename}{Supplementary Fig.}

\setcounter{equation}{0}
\setcounter{table}{0}

\setcounter{page}{1}

\newpage

\counterwithin*{figure}{part}
\counterwithin*{table}{part}

\stepcounter{part}

\begin{center}
\huge{\bf{Supplementary Material}}
\end{center}
\normalsize
\vspace{5 mm}

This document provides further information on: (\textit{i}) the correction of axial motion during each wavelength sweep; (\textit{ii}) the correction of axial motion between each plane wave illumination; (\textit{iii}) the theoretical expression of the holographic wave-field under a static assumption;  (\textit{iv}) the correction of transverse motion between each plane wave illumination.

\section{Correction of axial motion during each wavelength sweep} 
Figure~\ref{fig:gvd} illustrates the different steps of the group velocity dispersion correction originally proposed by Hillmann \textit{et al.}~\cite{Hillmann2012} and that constitutes the first step of our post-processing pipeline. The benefit of this axial motion correction can be appreciated by comparing the longitudinal section of a raw holographic image (Fig.~\ref{fig:gvd}b) with its corrected counterpart (Fig.~\ref{fig:gvd}h). The initial images are blurred due to eye axial motion during the wavelength sweep The effect of axial motion  causes the apparent optical path lengths in the sample arm to vary with the wavenumber. To circumvent that problem, Hillmann \textit{et al.}~\cite{Hillmann2012} proposed to perform a cross-correlation of sub-bandwidth images (Supplementary Fig.~\ref{fig:gvd}d) to determine the  depth shift $\Delta z_{\textrm{opt}}(\omega,\mathbf{u}_{\textrm{in}})$ (Eq.~\rev{17}, Fig.~\ref{fig:gvd}e) and the phase fluctuation $\phi(\omega,\mathbf{u}_{\textrm{in}})$ (Eq.~\rev{18}, Fig.~\ref{fig:gvd}f) due to axial motion. This phase error is compensated on each holographic field (Supplementary Fig.~\ref{fig:gvd}g) before obtaining the corrected images through an inverse Fourier transform along the frequency dimension (Fig.~\ref{fig:gvd}h). The photoreceptor layer that was fully blurred on the initial image (Fig.~\ref{fig:gvd}b) is now recovered with an excellent axial resolution (see white arrow in Fig.~\ref{fig:gvd}h). 

\begin{figure}
    \includegraphics[width=\textwidth]{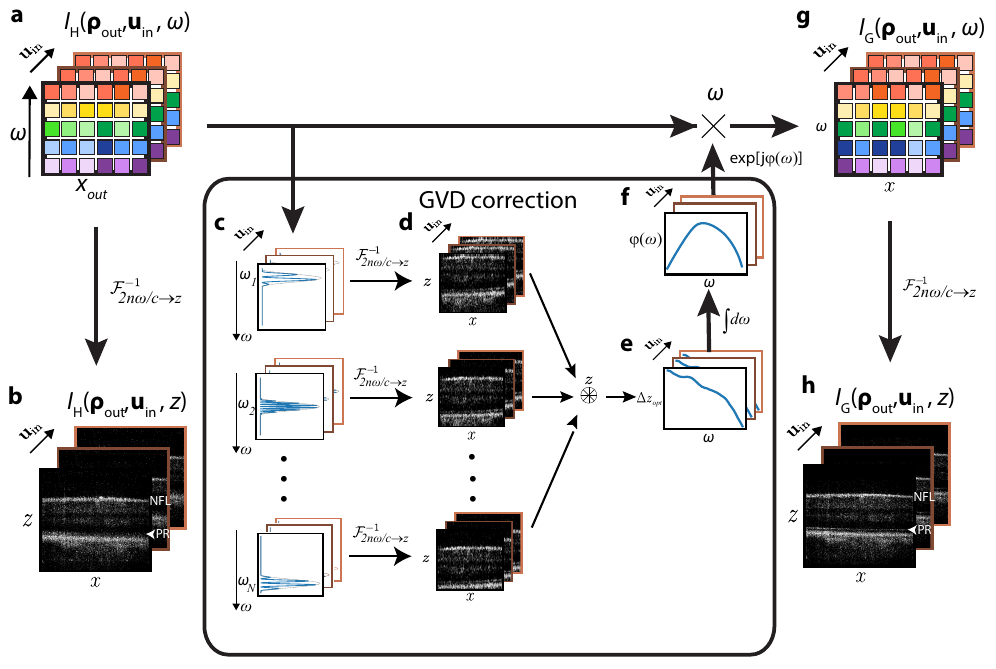}
    \caption{\textbf{Correction of axial motion during each wavelength sweep}. \textbf{a} Schematic representation of the multi-spectral holographic matrix $\mathbf{I}_H=[I_H(\bm{\rho_{\textrm{out}}},\mathbf{u}_{\textrm{in}},\omega) ]$. \textbf{b} Longitudinal section of images, $I_H(\bm{\rho_{\textrm{out}}},\mathbf{u}_{\textrm{in}},z)$ obtained by a direct Fourier transform over frequency $\omega$. Each image is blurred by eye axial motion during the wavelength sweep.  \textbf{c} Example of holographic signals, $I_F(\bm{\rho_{\textrm{out}}},\mathbf{u}_{\textrm{in}},\omega)$ for one camera pixel $\bm{\rho}_{\textrm{out}}$ after application of the sub-bandwidth filter $F(\omega-\omega_i)$. \textbf{d} Longitudinal section of corresponding filtered images $I_F(\bm{\rho_{\textrm{out}}},\mathbf{u}_{\textrm{in}},z)$ (Eq.~\rev{16}). \textbf{e}, An axial shift $\Delta z_{\textrm{opt}}(\omega,\mathbf{u}_{\textrm{in}})$ of the retina is estimated for each wavelength sweep by an axial cross-correlation of each image intensity (Eq.~\rev{17}). \textbf{f}, Phase function $\phi(\omega,\mathbf{u}_{\textrm{in}})$  obtained by integration of $\Delta z_{\textrm{opt}}(\omega,\mathbf{u}_{\textrm{in}})$  over the frequency bandwidth (Eq.~\rev{18}). \textbf{g}, Axial motion compensation is performed by applying $exp(-j \phi(\omega,\mathbf{u}_{\textrm{in}})$ to each hologram $I_H(\bm{\rho_{\textrm{out}}},\mathbf{u}_{\textrm{in}},\omega)$, yielding a corrected multi-spectral matrix  $\mathbf{I}_G=[I_G(\bm{\rho_{\textrm{out}}},\mathbf{u}_{\textrm{in}},\omega) ]$. \textbf{h}, Longitudinal section of corrected images, $I_G(\bm{\rho_{\textrm{out}}},\mathbf{u}_{\textrm{in}},z)$ obtained by a direct Fourier transform over frequency $\omega$.    }
    \label{fig:gvd}
\end{figure}

\clearpage

\section{Correction of axial motion between each plane wave illumination} 
Figure~\ref{fig:ax} illustrates the different steps of the axial motion compensation applied to dispersion corrected images, $I_G(\bm{\rho_{\textrm{out}}},\mathbf{u}_{\textrm{in}},z)$, between each plane wave illumination. Because axial motion during each wavelength sweep has been compensated for each plane wave illumination in an independent way, an axial shift $\Delta \rev{z} (\mathbf{u}_{\textrm{in}})$ subsists between each of those images (Fig.~\ref{fig:ax}a). Their incoherent compound is therefore far from being optimal with an axial blurring of each retina layer (Fig.~\ref{fig:ax}b). The en-face image at the NFL depth (Fig.~\ref{fig:ax}c) illustrates this feature by showing a superimposition of the NFL with a vessel sitting below it. Figure~\ref{fig:ax}d displays the axial shift $\Delta \rev{\hat{z}}(\mathbf{u}_{\textrm{in}})$ estimated across the illumination sequence (Eq.~\rev{19}). Figure~\ref{fig:ax}e shows the holographic images $I_R(\bm{\rho_{\textrm{out}}},\mathbf{u}_{\textrm{in}},z)$ after their axial shift by $ \Delta \rev{\hat{z}}(\mathbf{u}_{\textrm{in}})$ (Eq.~2). Their incoherent compound illustrates the benefit of this compensation by highlighting: (\textit{i}) a much better axial resolution of the different retina layers across its longitudinal section (Fig.~\ref{fig:ax}f) compared to the original image (Fig.~\ref{fig:ax}\rev{b}); a transverse section of the NFL layer (Fig.~\ref{fig:ax}g) no longer polluted by the vessel compared to its original counterpart (Fig.~\ref{fig:ax}c). 
\begin{figure}
    \includegraphics[width=\textwidth]{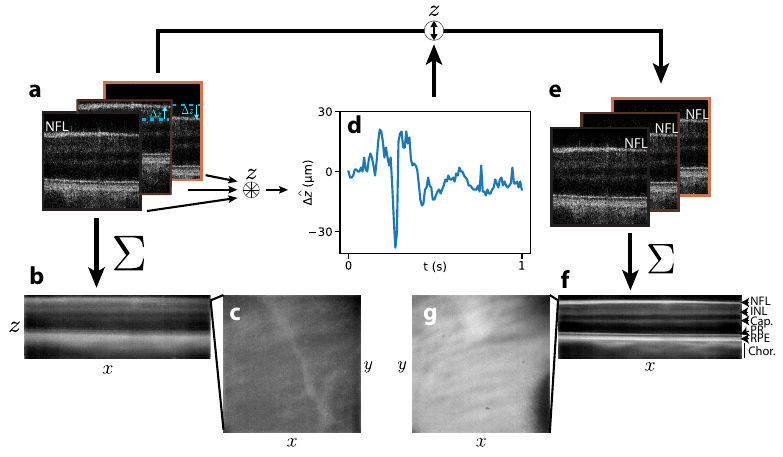}
    \caption{\textbf{Correction of axial motion between each plane wave illumination}.  \textbf{a} Longitudinal section of images, $I_G(\bm{\rho_{\textrm{out}}},\mathbf{u}_{\textrm{in}},z)$, obtained for each plan wave illumination after axial motion compensation during each wavelength sweep. Each image exhibits an axial shift $\Delta \rev{z}(\mathbf{u}_{\textrm{in}})$ with respect to each other. \textbf{b}-\textbf{c}, Longitudinal and transverse section of their incoherent compound, respectively. The en-face image (c) is displayed in the vicinity of the nerve fiber layer. \textbf{d}, Axial shift $\Delta \rev{\hat{z}}(\mathbf{u}_{\textrm{in}})$ estimated from the cross-correlation  of each image intensity with the reference image considered at $\mathbf{u}_{\textrm{in}}=\mathbf{0}$ (Eq.~18). \textbf{e}, Longitudinal section of images, $I_R(\bm{\rho_{\textrm{out}}},\mathbf{u}_{\textrm{in}},z)$, after axial motion compensation (Eq.~2) between each plan wave illumination $\mathbf{u}_{\textrm{in}}$. \textbf{f}-\textbf{g}, Longitudinal and transverse section of their incoherent compound, respectively. {The longitudinal image (f) now displays the different layers of the retina (NFL: nerve fiber layer; INL: Inner nuclear layer; Cap.: Capillaries; PR: Photoreceptors, RPE: Retina pigmental epithelium, Chor.: Choroidal tissues). The en-face image (g) is displayed at the depth of the nerve fiber layer.}}
    \label{fig:ax}
\end{figure}

\clearpage

\section{Expression of the interference wave-field under a static assumption} 
 In this part, we express theoretically the time-gated OCT signal, $I(\bm{\rho}_{\textrm{out}},\mathbf{u}_{\textrm{in}},z)$ under a static assumption. {This signal is obtained after a Fourier transform over frequency of the recorded holograms and corresponds to the cross-correlation part of the interference pattern isolated by an Heavy-side filter in the time domain~\cite{balondradeMultiSpectralReflectionMatrix2024}.} Under a single back-scattering approximation, this signal results from the interference between the wave-field, $E_S(\bm{\rho}_{\textrm{out}},\mathbf{u}_{\textrm{in}},z)$, back-scattered by the heterogeneities lying at depth $z$ inside the retina and the reference wave-field, $E_R(\bm{\rho}_{\textrm{out}},\mathbf{u}_{\textrm{in}},z)$, back-scattered by the mirror in the reference arm. 
\begin{equation}
\label{I0}
{I}(\bm{\rho}_{\textrm{out}},z,\mathbf{u}_{\textrm{in}})=E_S(\bm{\rho}_{\textrm{out}},\mathbf{u}_{\textrm{in}},z)E_R^*(\bm{\rho}_{\textrm{out}},\mathbf{u}_{\textrm{in}},z)
\end{equation}
 The retina wave-field $E_S(\bm{\rho}_{\textrm{out}},\mathbf{u}_{\textrm{in}},z)$ can be expressed as follows:
\begin{equation}
\label{ES}
{E}_S(\bm{\rho}_{\textrm{out}},z,\mathbf{u}_{\textrm{in}})=\int \int\, d\bm{\rho} h(\bm{\rho},\bm{\rho}_{\textrm{out}},z)  \gamma(\bm{\rho},z) E_{\textrm{in}}(\bm{\rho},\mathbf{u}_{\textrm{in}},z).
\end{equation}
where $E_{\textrm{in}}(\bm{\rho},\mathbf{u}_{\textrm{in}},z)$ is the incident wave-field inside the retina at depth $z$, $\gamma(\bm{\rho},z)$ is the retina reflectivity, and $h(\bm{\rho},\bm{\rho}_{\textrm{out}},z)$ is the impulse response of the imaging system between the camera plane ($\bm{\rho}_{\textrm{out}}$) and the retina at depth $z$. 
The incident wave-field can be expressed as:
\begin{equation}
\label{Einc}
E_{\textrm{in}}(\bm{\rho},\mathbf{u}_{\textrm{in}},z)=H(\bm{\rho},\mathbf{u}_{\textrm{in}},z)\exp \left ( - i\frac{2\pi}{\lambda f}\mathbf{u}_{\textrm{in}}.\bm{\rho} \right),
\end{equation}
The exponential term in the latter equation accounts for the Fourier transform operated by the eye lens between the pupil and its focal plane located at depth $z=0$. $f$ is the focal length of the eye lens.  $H(\bm{\rho},\mathbf{u}_{\textrm{in}},z)$ describes the transmittance between the pupil plane ($\mathbf{u}_{\textrm{in}}$) and each point $\bm{\rho}$ at depth $z$. The support of the transmittance in the pupil plane is dictated by the eye pupil size. Its phase accounts for the defocus undergone by incident light between depth $z$ and the focal plane ($z=0$) but also for the aberrations induced by the eye imperfections and refractive index heterogeneities inside the cornea or across retina layers.

Under an isoplanatic assumption, the transmittance $H(\bm{\rho},\mathbf{u}_{\textrm{in}},z)$ is invariant across the field-of-view:
\begin{equation}
\label{isoH}
H(\bm{\rho},\mathbf{u}_{\textrm{in}},z)=H(\mathbf{u}_{\textrm{in}},z)
\end{equation}
Similarly, the output spread function, $h$, is invariant by translation, such that
\begin{equation}
\label{isoh}
h(\bm{\rho},\bm{\rho}_{\textrm{out}},z)=h(\bm{\rho}-\bm{\rho}_{\textrm{out}},z)
\end{equation}
Note that both quantities are then related through a simple Fourier transform:
\begin{equation}
    H(\mathbf{u},z)=\int d\bm{\rho}' h(\bm{\rho}',z) \exp \left ( - i\frac{2\pi}{\lambda f}\mathbf{u}.\bm{\rho}' \right)
\end{equation}
Under the isoplanatic assumption (Eqs.~\ref{isoH}-\ref{isoh}) and injecting Eq.~\ref{Einc} into Eq.~\ref{ES} leads to the following expression for the wave-field $E_S$ backscattered by the retina:
\begin{equation}
\label{ES2}
{E}_S(\bm{\rho}_{\textrm{out}},z,\mathbf{u}_{\textrm{in}})=\int \int\, d\bm{\rho} H(\mathbf{u}_{\textrm{in}},z)  \gamma(\bm{\rho},z) h(\bm{\rho}-\bm{\rho}_{\textrm{out}},z)   \exp \left ( - i\frac{2\pi}{\lambda f}\mathbf{u}_{\textrm{in}}.\bm{\rho} \right).
\end{equation}
Under the same degree of approximation and assuming a reference mirror of unit reflectivity placed in the MO focal plane, the reference wave-field $E_R$ can be expressed as follows:
\begin{equation}
\label{ER}
{E}_{R}(\bm{\rho}_{\textrm{out}},z,\mathbf{u}_{\textrm{in}})=H_{\textrm{ref}}(\mathbf{u}_{\textrm{in}})   H_{\textrm{ref}}(-\mathbf{u}_{\textrm{in}})  \exp\left ( -i\frac{2\pi}{\lambda f}\mathbf{u}_{\textrm{in}}.\bm{\rho}_{\textrm{out}} \right).
\end{equation}
where $H_{\textrm{ref}}$ is the MO pupil function.
Injecting Eqs.~\ref{ES2} and \ref{ER} into Eq.~\ref{I0} leads to the following expression for the time-gated holographic signal $I$:
\begin{equation}
\label{I02}
{I}(\bm{\rho}_{\textrm{out}},z,\mathbf{u}_{\textrm{in}})= H_{\textrm{in}}(\mathbf{u}_{\textrm{in}},z)  \int \int\, d\bm{\rho}  \gamma(\bm{\rho},z) h(\bm{\rho}-\bm{\rho}_{\textrm{out}},z)   \exp \left [ - i\frac{2\pi}{\lambda f}\mathbf{u}_{\textrm{in}}.(\bm{\rho}-\bm{\rho}_{\textrm{out}}) \right].
\end{equation}
with $H_{\textrm{in}}$, the input aberration transmittance that accumulates the back-and-forth aberrations in the reference arm and the aberrations undergone by the incident light in the sample arm:
\begin{equation}
\label{Hin}
H_{\textrm{in}}(\mathbf{u}_{\textrm{in}},z)=H(\mathbf{u}_{\textrm{in}},z)H_{\textrm{ref}}(\mathbf{u}_{\textrm{in}})   H_{\textrm{ref}}(-\mathbf{u}_{\textrm{in}}). 
\end{equation}

\clearpage

\section{Correction of transverse motion between each plane wave illumination} 
Figure~\ref{fig:trans}b shows the incoherent compound of en-face images $\mathbf{I}_C(\mathbf{u}_{\textrm{in}},z)$ at the NFL depth (Fig.~\ref{fig:trans}a) obtained after axial motion and isoplanatic aberation compensation. This image is blurred by lateral motion of the retina during the illumination sequence. This drift prevents a proper combination of each image $\mathbf{I}_C(\mathbf{u}_{\textrm{in}},z)$. Lateral shift $\Delta{\bm{\rho}}(\mathbf{u}_{\textrm{in}})$ is estimated by cross-correlating each image, $I_C(\bm{\rho}_{\textrm{out}},\mathbf{u}_{\textrm{in}},z)$, with a reference one corresponding to normal plane wave illumination ($\mathbf{u}_{\textrm{in}}=\mathbf{0}$, Eq.~20). The estimated shift,  $\Delta \hat{\bm{\rho}}(\mathbf{u}_{\textrm{in}})=(\Delta \hat{x}(\mathbf{u}_{\textrm{in}},\Delta \hat{y}(\mathbf{u}_{\textrm{in}} )$, is displayed in Fig.~\ref{fig:trans}c. It shows an alternation between slow drift periods and micro-saccades characterized by abrupt retina motion. The estimated lateral shift $ \Delta \hat{\bm{\rho}}(\mathbf{u}_{\textrm{in}})$ is used to readjust each en-face image (Fig.~\ref{fig:trans}d). The incoherent compound of those images (Eq.~9) is shown in Fig.~\ref{fig:trans}e. The comparison with its initial counterpart (Fig.~\ref{fig:trans}b) demonstrates the validity of the transverse motion estimation and success of its compensation. The initial blurred feature is actually removed and a contrasted image of the NFL is recovered.   
\begin{figure}
    \includegraphics[width=\textwidth]{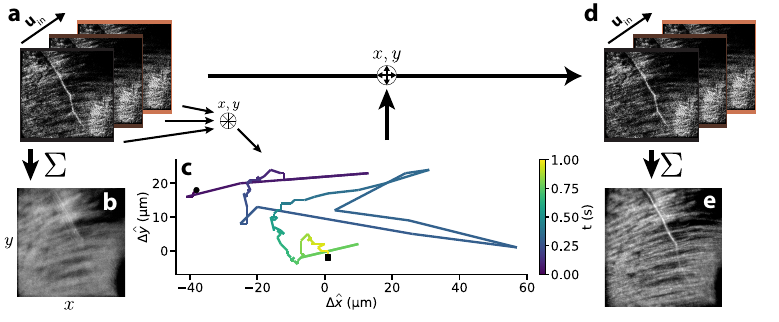}
    \caption{\textbf{Correction of transverse motion between each plane wave illumination}. \textbf{a} En-face image $I_C$ of the NFL for different plane wave illuminations $\mathbf{u}_{\textrm{in}}$ after axial motion and isoplanatic aberration compensation. \textbf{b} A direct incoherent compound of those images leads to a blurred image of the NFL  because of lateral motion. \textbf{c} Lateral shift estimation across the illumination sequence (Eq.~20). \textbf{d} Laterally shifted en-face image of the NFL (Eq.~8).  \textbf{e} The incoherent compound of those readjusted images (Eq.~9) leads to a contrasted image of the NFL. }
    \label{fig:trans}
\end{figure}

\end{document}